\documentclass{aa}
\usepackage{graphicx}
\usepackage{times}
\usepackage{longtable}
\newcommand{\SiIII}{\ion{Si}{iii}}
\newcommand{\MgII}{\ion{Mg}{ii}}
\newcommand{\HeI}{\ion{He}{i}}
\newcommand{\CLEAN}{{\sffamily\scshape clean\,}}
\newcommand{\KOREL}{{\sffamily\scshape korel\,}}
\newcommand{\FOTEL}{{\sffamily\scshape fotel\,}}
\newcommand{\XSPEC}{{\sffamily\scshape xspec}}
\begin{document}
 \title{The orbit of the close spectroscopic binary \object{$\varepsilon\,$Lup} and the intrinsic variability of its early B-type components
\thanks{Based on spectral observations obtained at ESO with FEROS/2.2m,
La~Silla, Chile and at SAAO with GIRAFFE/1.9m, Sutherland, South Africa}
\fnmsep\thanks{Table~\ref{onlinedata} is only available in electronic form at http://www.edpsciences.org}
}
\author{K.\ Uytterhoeven\inst{1,2} \and P.~Harmanec\inst{2,3} \and
J.H. Telting\inst{4} \and C.\ Aerts\inst{1,5}}
  \offprints{K.\ Uytterhoeven: \\ katrienu@sunstel.asu.cas.cz}
  \institute {
   Institute of Astronomy, Katholieke Universiteit Leuven,
   Celestijnenlaan~200~B, B-3001~Leuven, Belgium
\and
   Astronomical Institute, Academy of Sciences, Fri\v{c}ova 298,
   CZ-251~65~Ond\v{r}ejov, Czech Republic
\and
   Astronomical Institute of the Charles University,
   V Hole\v{s}ovi\v{c}k\'ach~2, CZ-180~00~Praha 8, Czech Republic
\and
Nordic Optical Telescope, Apartado de Correos 474, E-38700 Santa Cruz de La Palma, Spain
\and
Department of
    Astrophysics , University of Nijmegen, PO Box 9010, NL-6500 GL
   Nijmegen, The Netherlands
}

\authorrunning{K.\ Uytterhoeven et al.}
\titlerunning{The close binary system $\varepsilon\,$Lup}

\date{Received; Accepted}

\abstract{We subjected 106 new high-resolution spectra of the double-lined
spectroscopic close binary $\varepsilon\,$Lup, obtained in a time-span of
17 days from two different
observatories, to a detailed study of
orbital and intrinsic variations.  
We derived accurate values of the orbital parameters. We refined the sidereal orbital period to $4.\!\!^{\rm d}55970$ days
and the eccentricity to $e=0.277$. By adding old radial velocities, we discovered the presence of apsidal motion with a period of the rotation of apses of about 430 years. Such a value  agrees with theoretical expectations. Additional data is needed to confirm and refine this value. Our dataset did not allow us to derive the orbit  of the third body, which is known to
orbit the close system in $\sim$ 64 years.
We present the secondary of $\varepsilon\,$Lup as a new $\beta$~Cephei\,
variable, while the primary is a $\beta$~Cephei\, suspect.
A first detailed analysis of line-profile variations of both primary
and secondary led to detection of one pulsation frequency near
10.36 c d$^{-1}$ in the variability of the secondary, while no clear
periodicity was found in the primary, although low-amplitude
periodicities are still suspected. The limited accuracy and extent of our dataset did not
allow any further analysis, such as mode-identification.  } \maketitle

\section{Introduction}
Line-profile variable early B-type stars which belong to a close binary system are very interesting targets for  studying the effect of tidal interactions on pulsation-mode selection and/or on the enhancement of pulsation-mode amplitudes.  
As the number of cases observed with confirmed tidally induced modes
is sparse\, (HD\,177\,863, De Cat \& Aerts 2002; HD\,209\,295, Handler
et al.\ 2002), two complementary surveys have been set up to 
systematically study the behaviour of non-radial pulsations (NRPs) in early B-type stars with a close companion. The first survey is called SEFONO
(SEarch for FOrced Non-radial Oscillations, Harmanec et al.\ 1997), which searches for line-profile variations (LPVs) in the early-type primaries of known,
short-period binary systems and studies orbital forcing as a possible
source of NRPs. The second survey has a complementary approach (Aerts et al.\ 1998) and starts from a sample of selected main-sequence
early-type B stars without Balmer line emission, which are known to exhibit NRPs and which turn out to be
part of a close binary.
 As high-resolution spectrographs allow
detailed study of low-amplitude features moving through the line
profiles, both surveys are spectroscopic surveys and focus on the
presence of LPVs as signatures of the forced oscillation modes. Aerts \& Harmanec (2004) have compiled a catalogue of line-profile variables in close binaries.

In the framework of such a systematic study of line-profile variable early B-type stars, we selected known close binary systems from the list of candidate $\beta$ Cephei variables resulting from the high-resolution spectroscopic survey for LPV in 09.5-B2.5 II-V stars performed by Schrijvers et al.\ (2002) and Telting et al.\ (2003).  One of the interesting results of this survey is detection of LPVs in 16 of the 27 known short-period binaries ($P_{\rm orb} < 10^{\rm d}$) in the sample (Telting et al.\, in preparation).
 In this paper we present results from the study of the eccentric binary $\varepsilon\,$Lup.

\section{Present knowledge about $\varepsilon\,$Lup}
\begin{figure}
\resizebox{0.45\textwidth}{!}{\includegraphics{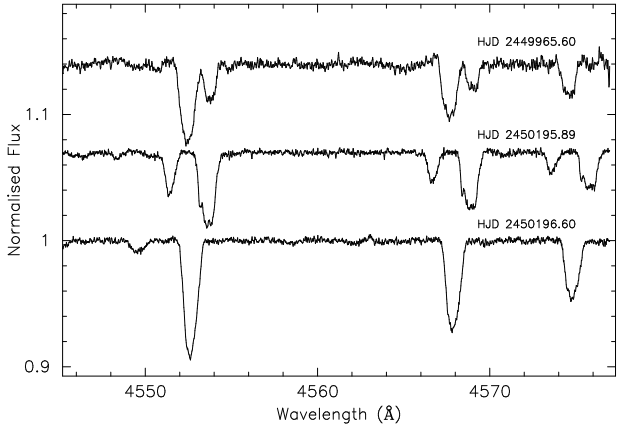}}
\caption{Spectra of $\varepsilon\,$Lup
centered at the \SiIII\ triplet near 4568~\AA\ 
obtained during nights in September 1995 and April 1996  with the ESO CAT/CES combination (Schrijvers et al.\ 2002). The spectra are offset for
clarity.}
\label{CATspectra}
\end{figure}

The target $\varepsilon\,$Lup (\object{HD\,136\,504}, HR\,5708, $\alpha_{2000}=15^{\rm h}\,22^{\rm m}\,40.\!\!^{\rm s}87$, $\delta_{2000}=-44^{\circ}\,41^{\prime}\,22.\!\!''64$, $m_V=3.367$) is a double-lined spectroscopic binary (SB2) in a close orbit (Curtis 1909; Moore 1910; Campbell
\& Moore 1928; Buscombe \& Morris 1960; Buscombe \& Kennedy 1962; Thackeray 1970)
 and  the brighter component of a visual binary  with a period of
several decades (Thackeray 1970, hereafter T70). Buscombe \& Kennedy  (1962)  
published the first orbital solution, resulting in a 
nearly circular orbit ($e\leq0.03$) with an orbital period of
approximately $0.\!\!^{\rm d}901$ 
days. Regrettably, their radial velocities (RVs) were never published. T70 determined an orbital
period of $4.\!\!^{\rm d}5597$ days and an eccentricity of $e = 0.26$.  No signs of
cyclic changes due to ellipticity were detected, and no eclipses were
observed. T70 reported the presence of the
third component (separation $26''.5$) with a possible orbital motion
of 64 years. T70 provided spectral types for all
three  components, B3IV, B3V, and most likely A5V. Other estimates found in the literature provide only one type for the whole multiple system, which ranges from B3IV to B2IV-V (de Vaucouleurs 1957; Levato \& Malaroda 1970; Hiltner et al. 1969; BSC, Hoffleit 1991). No other studies of the orbital parameters  have been carried out so far.

   The HIPPARCOS parallax resulted in a distance of  $155\pm15$\,pc 
(Perryman et al.\ 1997), which led De Zeeuw et al.\ (1999) to
conclude that $\varepsilon\,$Lup is most probably a member of the UCL OB association, located at a
distance of 140 pc. In this case, the estimated age of the multiple
system of $\varepsilon\,$Lup is about 14-15 Myr.  

 Tokovinin (1997) gives the following values of the visual magnitudes
of the separate components of the system AaB: $m_{V_{Aa}} = 3.62$,  $m_{V_{B}} = 5.10$.  
In his catalogue we also find estimates for the masses, based on the mass function of the orbital solution presented by T70: $M_{B} = 7.64
M_{\odot}$,  $M_{Aa} = 24.70 M_{\odot}$, whereby $M_{A} = 13.24 M_{\odot}$ and
$M_{a} = 11.46 M_{\odot}$. Based on Geneva photometry, De Cat (2002) lists $M = 7.1 \pm 0.6
M_{\odot}$ and $R = 5.0 \pm 1.0 R_{\odot}$, assuming a single star.
The following estimates of the effective temperature,
$\log g$, and the luminosity are available in the literature:
$T_{\rm eff} = 21\,230$ K (visible spectrophotometry, Morossi \& Malagnini 1985); $T_{\rm eff} =
23\,410 \pm 2\,510$ K (spectroscopy, Sokolov 1995); $T_{\rm eff} = 19\,240$ K, $\log g =
3.85$ ($ubvy$ photometry, Castelli 1991); $T_{\rm eff} = 19\,275 \pm 1800$ K, $\log g =
3.89 \pm 0.17$, $\log L = 3.50 \pm 0.18 L_{\odot}$ (Geneva photometry, De Cat 2002).
 
In the literature we find different values of the projected rotational
velocity $v \sin i$: 166~km~s$^{-1}$ (Bernacca \& Perinotto 1970); 142 
km s$^{-1}$ (Uesugi \& Fukuda 1970); 170 km s$^{-1}$ (Levato \&
Malaroda 1970); 40 km s$^{-1}$ (Slettebak et al. 1975); 133 km s$^{-1}$ (Hoffleit 1991). The higher values are obtained by treating the
system as a single star.

The intrinsic variability of $\varepsilon\,$Lup has not yet been studied in detail. 
Schrijvers et al.\ (2002) detected bumps in the primary,
possibly caused by NRP of high-degree ($\ell > 2$). Their spectra, taken with the CAT/CES combination at ESO, La Silla, during  nights in September 1995 and April 1996,  are given in Fig.~\ref{CATspectra} and are included in the analysis presented in this paper. Signatures of both duplicity and line-profile variability of the primary are visible.
Both components of the spectroscopic binary are located in the
theoretical $\beta$~Cephei\, instability strip
(Pamyatnykh 1999) making them candidates for exhibiting $\beta$~Cephei\, type pulsations. 

\section{Data and data reduction}
\begin{table}
\caption{Journal of the observations of $\varepsilon\,$Lup obtained in
May 2003. The top part is the logbook of the 56 FEROS (11--16 May 2003), while the bottom part gives information on the 47
observations made with GIRAFFE (20--26 May 2003). The columns
list the Julian date of the observations, the number of
spectra obtained, the mean S/N-ratio, the mean integration time expressed in
seconds, and the number of spectra for which  the profiles of primary
and secondary are separated.}
\label{logepslup}
\begin{flushleft}
\begin{tabular}{lllcc}
\noalign{\smallskip}\hline
\hline\noalign{\smallskip}
Date (HJD) & N & S/N & $\Delta t$ (s) & N$_{\rm separate}$\\ \hline
2452770 & 18  & 450 & 330 & 0 \\
2452771 & 13  & 550 & 390 & 13 \\
2452772 & 6   & 460 & 490 & 0 \\
2452773 & 10  & 570 & 405 & 10 \\
2452774 & 1   & 250 & 600 & 0 \\
2452775 & 8   & 400 & 420 & 0 \\ \hline
2452780 & 6   & 190 & 490 & 3 \\ 
2452781 & 13  & 210 & 370 & 10 \\
2452783 & 16  & 200 & 450 & 11 \\
2452784 & 12  & 185 & 365 & 0 \\ 
\noalign{\smallskip}\hline
\end{tabular}
\end{flushleft}
\end{table}

\begin{table*}
\caption{Journal of the available RVs of the primary and secondary.}
\label{RVjournal}
\begin{flushleft}
\begin{tabular}{llccccl} \noalign{\smallskip}\hline
\hline\noalign{\smallskip}
Nr. & Observatory & Telescope/Instrument & Time interval& $\sharp$ RVs& $\sharp$  RVs & Dispersion\\
 &             &            & (HJD$-$2400000) & comp1  & comp2 & (\AA/mm$^{-1}$) \\ \hline
1 & Lick Southern Station & 0.929m/1-\& 2-prism spectrograph & 17691.76-19221.59 & 10 & 7  & 10.3-20.3   \\
2 & Mount Stromlo         & 0.762m/3-prism spectrograph       & 35297.96-36798.86 & 3  & 3  & 36 \\
3 & Radcliffe             & 1.9m/2-prism spectrograph       & 38456.59-40369.36 & 32 & 18 & 6.8-15.5\\
4 & ESO                   & CAT/CES                    & 49965.60-50196.60 & 3  & 3  & R=65000 \\
5 & ESO                   & 2.2m/FEROS                 & 52770.51-52775.86 & 56 & 25 & R=48000 \\
6 & SAAO                  & 1.9m/GIRAFFE               & 52780.28-52784.61 & 47 & 39 & R=32000 \\ \noalign{\smallskip}\hline
\multicolumn{4}{l}{\scriptsize{1: Campbell \& Moore (1928); 2: Buscombe \& Morris (1960); 3: T70; 4: Schrijvers et al.\ (2002)}}
\end{tabular}
\end{flushleft}
\end{table*}

 We obtained  two quasi-consecutive time-series of high-resolution \'echelle spectra for $\varepsilon\,$Lup, obtained from different observatories, in order to cover the orbit as well as possible.

\begin{figure}
\begin{center}
\resizebox{0.40\textwidth}{!}{\includegraphics{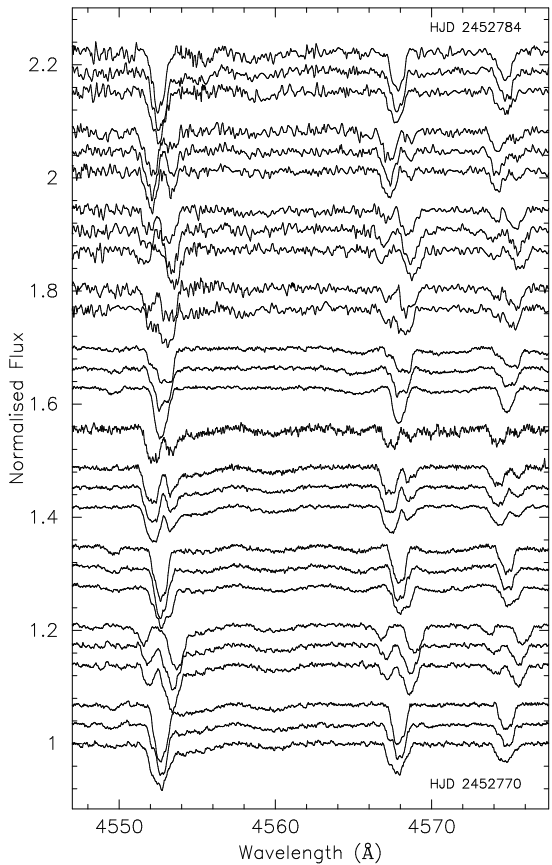}}\caption{A randomly chosen set of reduced spectra of $\varepsilon\,$Lup
centered at the \SiIII\ triplet near 4568~\AA\ 
obtained during respectively 6 and 4 nights in May 2003 with the FEROS 
(bottom) and GIRAFFE (top) spectrograph. The spectra are offset for
clarity. }
\label{epslupspectra}
\end{center}
\end{figure}

The first part of the time-series of $\varepsilon\,$Lup was gathered with the
 FEROS \'echelle spectrograph attached to the 2.2m ESO telescope at
 ESO, La Silla, Chile during the nights of 11--16 May 2003. The night
 of 15 May was mainly lost due to bad weather. The spectra have a resolution $R\sim 48000$ and cover the range 3500 to 9200~\AA, divided  into 39 orders. 
We obtained 56 spectra with an average S/N-ratio of 450 in the
 spectral range of the \SiIII\ triplet near 4568 \AA. The integration
 times were between 5 and 13 minutes. A logbook of the 6 nights is
 given in the top part of Table~\ref{logepslup}. 
The data reduction was performed using the on-line FEROS reduction pipeline,  which makes
 use of the ESO-MIDAS software package. We performed an additional correction
 for the wavelength sensitivity  of the shape of the internal
 flatfields  by means of a smoothed average of dome flatfields. 
  For each of the 6 nights, a randomly chosen set of
 normalised FEROS spectra centered at the \SiIII\ triplet near 4568~\AA\ is given at the bottom of Fig.~\ref{epslupspectra}.

The second part of the time-series of $\varepsilon\,$Lup consists of 47
spectra ($R \sim 32\,000$) measured with the GIRAFFE \'echelle spectrograph at the
1.9m telescope at SAAO, Sutherland, South Africa. From the 7 nights of
20--26 May 2003, 3 nights were lost due to bad weather. Integration
times were between 5 and 10 minutes. 
 The total wavelength range covers 4400-6680~\AA\  and
is spread over 45 orders. Due to the lower resolution the GIRAFFE
data are of lower quality than the FEROS data. The average S/N-ratio we
obtained was 200. A logbook of the spectra can be found in the lower
part of  Table~\ref{logepslup}. 
The GIRAFFE spectra were reduced by using the GIRAFFE pipeline reduction
program \XSPEC2. 
At the top of Fig.~\ref{epslupspectra}  a sample of reduced and normalised
GIRAFFE spectra centered on the wavelength region of the \SiIII\ triplet near 4568 \AA\ are shown.

\section{The orbital motion}
\begin{table*}
\caption{Orbital elements and their standard errors of the close
binary system of $\varepsilon\,$Lup, derived with the \FOTEL code from the RVs of the individual dataset of Lick (second column) and Radcliffe (third column) data, of the combined Lick \& Radcliffe data (fourth column), and of FEROS \& GIRAFFE data (last column).  Weights according to the formula $w = R/32000$ were assigned and  data points with errors larger than 10 km s$^{-1}$ were removed.  All epochs are given in HJD-2400000.  As comparison, the first column gives the solution of the system, as presented by T70, which is  based on 18 Radcliffe spectra.   The dataset numbers are as indicated in Table~\ref{RVjournal}.}
\label{epsluporbsolution1}
\begin{flushleft}
\begin{tabular}{llllll}
\noalign{\smallskip}\hline
\hline\noalign{\smallskip}\tabcolsep=1pt  
Nr. &   3        & 1      & 3      & 1+3    & 5+6     \\
Elem. & T70  & \FOTEL & \FOTEL & \FOTEL & \FOTEL  \\ \hline
$P_{\rm orb}$ (days) & $4.559783 \pm 2~10^{-6}$ & $4.5597$ \small{fixed} & $4.5597 \pm 1~10^{-4}$ &$4.55977 \pm 1~10^{-5}$ & $4.560 \pm 8~10^{-3}$  \\
$T_0$  & $39370.68 \pm 0.09$ & $18304.4 \pm 0.3$ & $39370.71 \pm 0.08$ & $39370.68 \pm 0.08$ & $52767.58 \pm 0.01$ \\
$e$ & $0.26 \pm 0.03$ & $0.26$ \small{fixed} & $0.26 \pm 0.03$& $0.26 \pm 0.03$& $0.300 \pm 0.006$  \\
$K_1$ (km s$^{-1}$)& $56.1 \pm 1.5$ &$55 \pm 5$ &$51 \pm 2$ &$52 \pm 2$&$54.0 \pm 0.3$ \\
$K_2$ (km s$^{-1}$)& $64.8 \pm 1.8$ &$71 \pm 5$ &$69 \pm 2$&$69 \pm 2$& $64.0 \pm 0.3$ \\
$\omega$ ($^{\circ}$)& $330 \pm 10$ & $319 \pm 25$& $332 \pm 8$&$329 \pm 7$& $24 \pm 1$ \\
$a_1\sin{i}$ (A.U.)& $0.0227$ & $0.022 \pm 2~10^{-3}$ & $0.0206 \pm 8~10^{-4}$ & $0.0210\pm 8~10^{-4}$ & $0.0215 \pm  5~10^{-4}$ \\
$a_2\sin{i}$ (A.U.)& $0.0262$ & $0.0289 \pm 2~10^{-3}$ & $0.0279 \pm 8~10^{-4}$ & $0.0278 \pm 8~10^{-4}$ & $0.026 \pm 5~10^{-3}$ \\
$f(M_2)$ ($M_{\odot}$)  & $0.06 \pm 0.01$ & $0.07 \pm 0.02$ & $0.056 \pm 0.007$& $0.060 \pm 0.007$& $0.065 \pm 0.005$\\
$f(M_1)$ ($M_{\odot}$)  & $0.14 \pm 0.02$& $0.16 \pm 0.03$& $0.14 \pm 0.01$ &$0.14 \pm 0.01$ & $0.107 \pm 0.007$ \\
$v_{\gamma_{\rm j}}$ (km s$^{-1}$)& $8 \pm 1$ & {\tiny j=1:} $5 \pm 3$ & {\tiny j=3:} $9 \pm 1$ & {\tiny j=1:} $6 \pm 3$ & {\tiny j=5:} $ 3.2 \pm 0.3$  \\
                            & --- & --- & --- & {\tiny j=3:} $8 \pm 1$ & {\tiny j=6:} $ 0.4 \pm 0.3$  \\ 
rms$_{\rm j}$  (km s$^{-1}$)& --- & {\tiny j=1:} $ 8.8$ & {\tiny j=3:} $7.6$ & {\tiny j=1:} $9.0$ & {\tiny j=5:} $2.6$   \\
                      & --- & --- & --- & {\tiny j=3:} $7.6$ & {\tiny j=6:}  $2.1$ \\
rms      (km s$^{-1}$)& 3.6 & 8.8 & 7.6 & 8.1 & 2.4  \\ 
\hline
$P_{\rm orb,wide}$ (yrs)& 64 & ---& ---& ---& ---  \\ 
\noalign{\smallskip}\hline
\end{tabular}
\end{flushleft}
\end{table*}

T70 studied the orbital motion of the triple system
$\varepsilon\,$Lup from spectra obtained with the Radcliffe
telescope during 1964-1970. The orbital elements
he derived  are listed in the first column of
Table~\ref{epsluporbsolution1}. 
Our dataset, consisting of 103 spectra obtained from two different observatories, samples
the short orbital period approximately 3 times, whereby 80\% of all phases is
covered. Our short dataset, in combination with the older data, does not allow  investigation of the triple  system, due to the extremely poor phase
coverage. 

\subsection{Radial velocities}
\label{apsmotion}

\begin{figure}
\begin{center}
\begin{tabular}{c}
\resizebox{0.45\textwidth}{!}{\rotatebox{-90}{\includegraphics{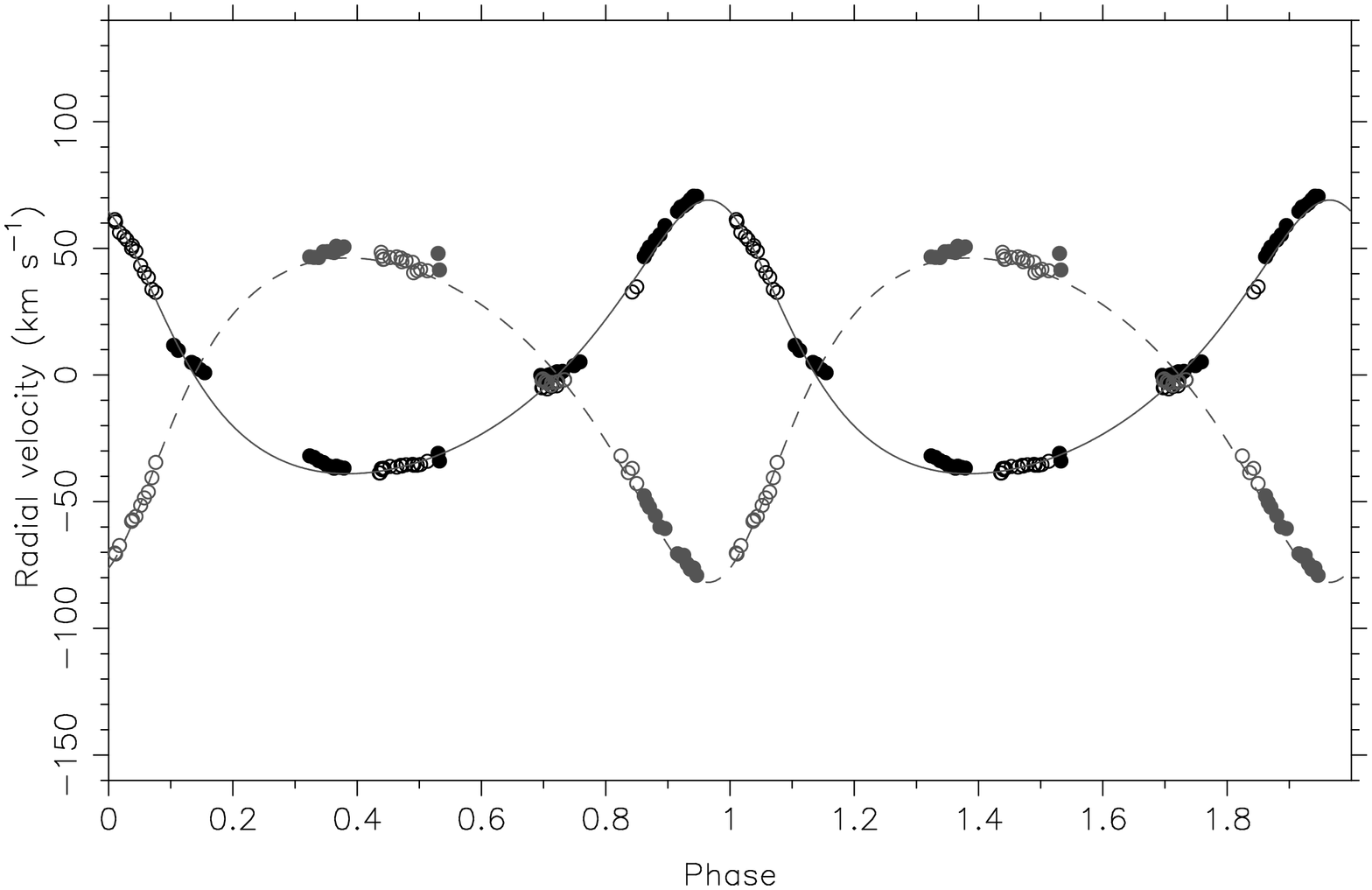}}} \\
\resizebox{0.45\textwidth}{!}{\rotatebox{-90}{\includegraphics{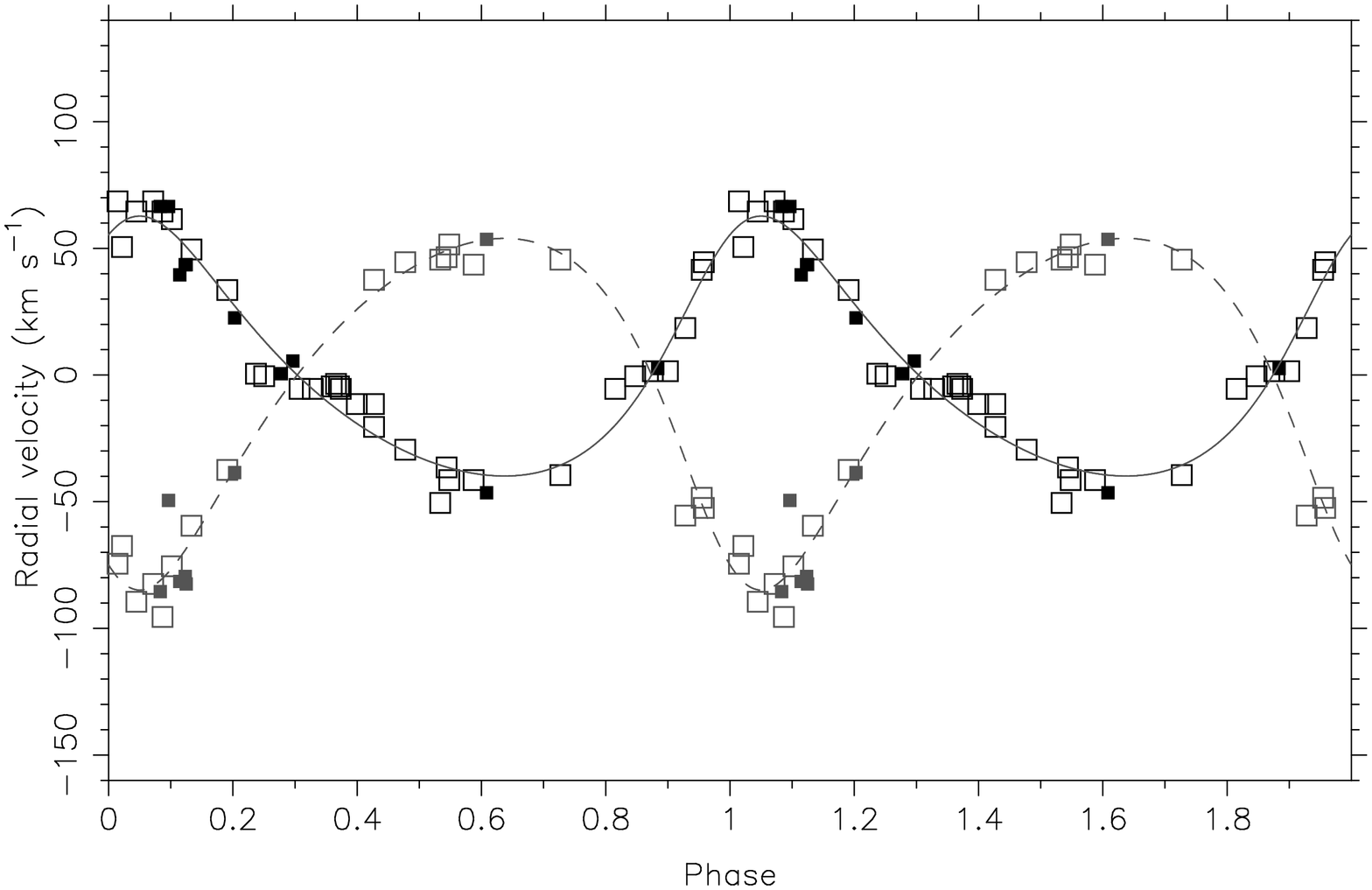}}} \\
\resizebox{0.45\textwidth}{!}{\rotatebox{-90}{\includegraphics{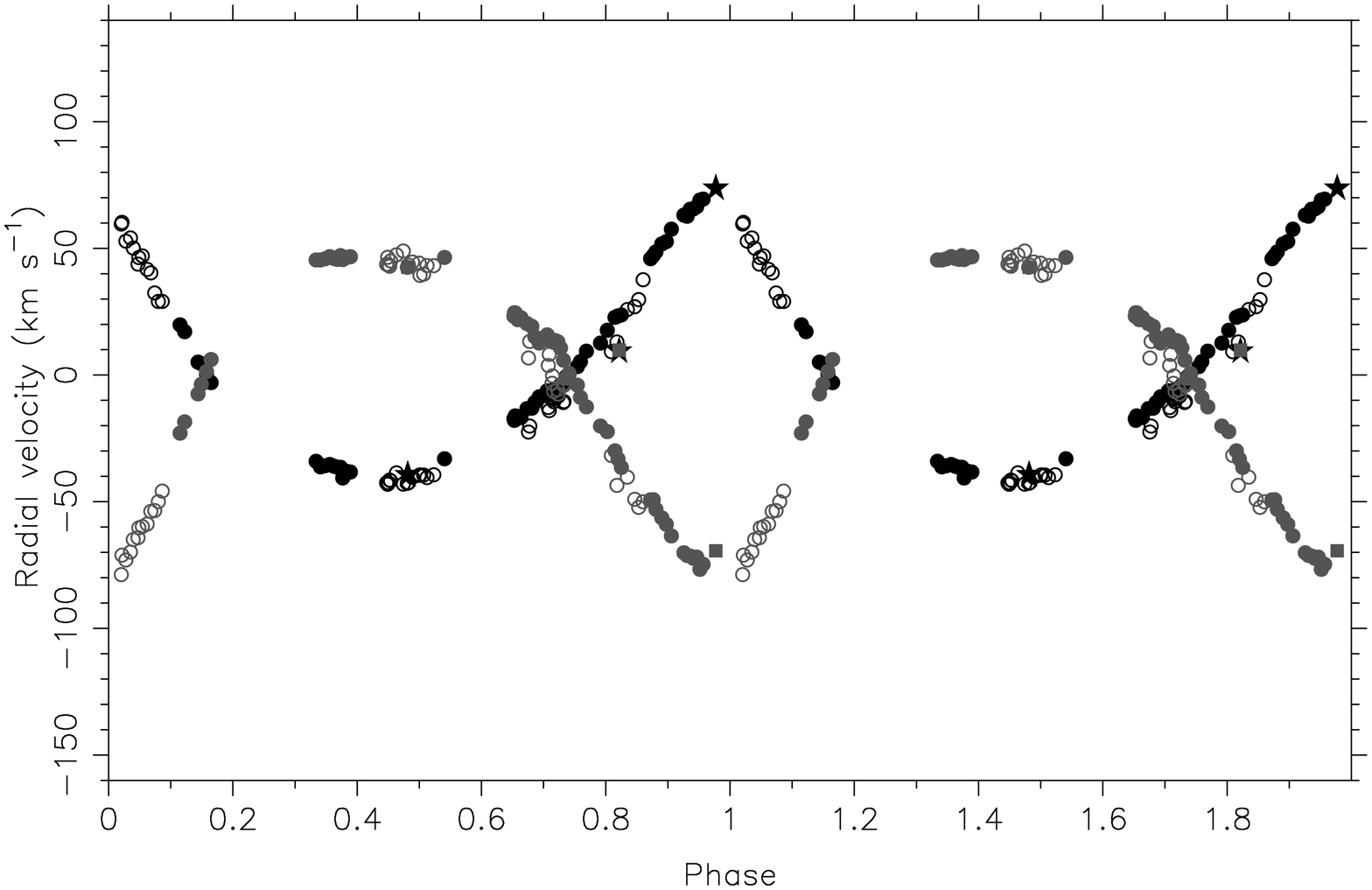}}} \\
\end{tabular}
\caption{Observed RVs of $\varepsilon\,$Lup derived from the FEROS ($\bullet$) and GIRAFFE ($\circ$) data (top) and older Lick ({\tiny black square}) and Radcliffe ({\tiny open square}) data (middle) versus orbital phase ($P_{\rm orb} = 4.\!\!^{\rm d}560$ and $P_{\rm orb} = 4.\!\!^{\rm d}55977$, respectively). The dark gray full and dashed lines represent
the local orbital solutions obtained with \FOTEL 
(last two of columns Table~\ref{epsluporbsolution1}) for the primary (black) and
secondary (gray), respectively. Note the different shapes of the RV curves, reflecting the slow apsidal advance.
In the bottom figure the RVs calculated from the CAT ($\star$), FEROS ($\bullet$), and GIRAFFE
($\circ$) spectra by \KOREL spectral disentangling, including $\dot{\omega}$ into the solution, versus orbital phase ($P_{\rm orb} = 4.\!\!^{\rm d}55970$, last column Table~\ref{epsluporbsolution2}). Phase zero corresponds to
the periastron passage.}
\label{epslupfotelplot}
\end{center}
\end{figure}

\begin{table*}
\begin{flushleft}
\begin{minipage}[l]{0.58\linewidth}
\begin{tabular}{llll}
\noalign{\smallskip}\hline
\hline\noalign{\smallskip}
Nr. & 1,3-6  & 3-6 & 4,5,6 \\
Elem. & \FOTEL & \FOTEL & \KOREL \\ \hline
$P_{\rm orb,ano}$ (days) & $4.55983 \pm 1~10^{-5}$ & $4.5598 \pm 3~10^{-4}$ & $4.55983$ \small{fixed} \\
$P_{\rm orb,sid}$ (days) & $4.55970 \pm 1~10^{-5}$ & $4.5597 \pm 3~10^{-4}$ & $4.55970$ \small{fixed} \\
$T_0$  & $39370.78 \pm 0.05$ & $52790.35 \pm 0.02$ & $52790.33$ \\
$e$ & $0.272 \pm 0.006$ & $0.271 \pm 0.005$ & $0.277$ \\
$K_1$ (km s$^{-1}$)& $53.9 \pm 0.4$ & $53.9 \pm 0.3$ & $53.8$ \\
$K_2$ (km s$^{-1}$)& $64.0 \pm 0.4$ & $64.0 \pm 0.3$ &$64.7$ \\
$\omega$ ($^{\circ}$)& $347 \pm 5$ & $18 \pm 2$ & $17$ \\
$\dot{\omega}$ ($^{\circ}$/yr)& $0.8 \pm 0.2$ & $1.1 \pm 0.2$ & $0.8$ \small{fixed}\\
$U$  (years)& $428 \pm 79$ & $319 \pm 55$ & $428$ \small{fixed} \\
$a_1\sin{i}$ (A.U.)& $0.0217 \pm 2~10^{-4}$ &$0.0217 \pm 1~10^{-4}$ & $0.0216$\\
$a_2\sin{i}$ (A.U.)& $0.0258 \pm 2~10^{-4}$ &$0.0258 \pm 1~10^{-4}$ & $0.0261$\\
$f(M_2)$ ($M_{\odot}$)& $0.066 \pm 0.002$& $0.066 \pm 0.001$&$0.065$\\
$f(M_1)$ ($M_{\odot}$)& $0.110 \pm 0.002$& $0.110 \pm 0.002$&$0.114$\\
$v_{\gamma_1}$ (km s$^{-1}$)& $4 \pm 3$ & --- & --- \\ 
$v_{\gamma_3}$ (km s$^{-1}$)& $9 \pm 2$ & $9 \pm 2$ &--- \\
$v_{\gamma_4}$ (km s$^{-1}$)& $3 \pm 3$ & $3 \pm 3$ &--- \\
$v_{\gamma_5}$ (km s$^{-1}$)& $1.1 \pm 0.2$ & $1.1 \pm 0.2$ & --- \\
$v_{\gamma_6}$ (km s$^{-1}$)& $-0.3 \pm 0.3$ & $-3.0 \pm 0.2$ &--- \\
rms$_1$ (km s$^{-1}$)& 11.2 & ---  & --- \\
rms$_3$ (km s$^{-1}$)& 9.8  & 8.2  & --- \\
rms$_4$ (km s$^{-1}$)& 5.8  & 5.3  & --- \\  
rms$_5$ (km s$^{-1}$)& 1.9  & 1.8  & --- \\ 
rms$_6$ (km s$^{-1}$)& 2.2  & 2.2  & --- \\
rms     (km s$^{-1}$)& 2.9  & 2.6  & --- \\  
\noalign{\smallskip}\hline
\end{tabular}
\end{minipage}
\hfill
\begin{minipage}[l]{0.36\linewidth}
\caption{Orbital elements and their standard errors for the close
binary system of $\varepsilon\,$Lup, derived with the \FOTEL code from all RV data, except Mt Stromlo (first column), and from the subset of  Radcliffe, CAT, FEROS, and GIRAFFE data (second column), including apsidal advance. In the presented solutions, the RVs of CAT, FEROS, and GIRAFFE were replaced by the corresponding \KOREL RVs.  Weights according to the formula $w = R/32000$ were assigned and  data points with errors larger than 10 km s$^{-1}$ were removed. The last column gives the results associated to the
best \KOREL disentangling solution of CAT, FEROS, and GIRAFFE spectra, in which the values of $P_{\rm orb}$ and $\dot{\omega}$ were kept fixed. Both  anomalistic and sidereal period are listed. All epochs are given in HJD-2\,400\,000.  The dataset numbers correspond to the ones  in Table~\ref{RVjournal}.}
\label{epsluporbsolution2}
\end{minipage}
\end{flushleft}
\end{table*}

We derived the RV values of each of the two components in
two different  ways: from the first normalised velocity moment $\langle v\rangle$ (e.g.\ Aerts et al.\ 1992) 
and by means of the cross-correlation (CC) technique. 

To calculate of the $\langle v\rangle$ of the three lines of the \SiIII\ triplet
at 4552.654~\AA, 4567.872~\AA\ and 4574.777~\AA, we used variable integration boundaries. Values
 of the RVs of the secondary could only be
derived at phases near  elongation, when both profiles were separated. One has to keep in mind that the estimates of the RVs of the primary near conjunction are contaminated by the presence of the secondary.

We applied the CC technique to several wavelength regions with well-defined absorption lines for primary and secondary. These
regions were centered on the following 6 lines: \SiIII~4568~\AA, 
\MgII~4481~\AA, \HeI~5016~\AA,  \HeI~4917~\AA, \HeI~5876~\AA, and \HeI~6678~\AA.  As templates for the CC, we took the
average normalised spectrum of the FEROS and of the GIRAFFE datasets.
The position of the line center, measured by means of a Gaussian fit, was used as an estimate of the RV.
The CC technique also allowed us to determine
the RVs of the secondary when its
profiles were semi-detached from those of the primary.  
Finally, we calculated the median RV of the 6 regions and its standard deviation.

It turned out that the scatter on the first velocity moment
calculated from the \SiIII\ profiles of the FEROS spectra was less
than for the RVs calculated by CC. This is possibly due to
the good quality and high-resolution of the FEROS spectra and the
nicely defined absorption profiles of the \SiIII\ triplet. Therefore
we used the values of the average of the  first moments of the 4553
\AA\ and 4568~\AA\ profiles  as RVs of the FEROS time-series in the
subsequent analysis.  
The GIRAFFE spectra, on the other hand,  are  of lower quality than the FEROS spectra. For these spectra the noise was suppressed by the CC procedure, so that we used the median of the CC measurements as RVs of the  GIRAFFE time-series. 

Additionally, we included  the three high-quality (R$\sim$65\,000, S/N $>$ 450) CAT spectra, obtained by Schrijvers et al.\ (2002) in 1995-1996. We calculated the first normalised velocity moment  $\langle v\rangle$ for both primary and secondary. 

We enlarged our sample of RV measurements with the ones T70 used for his analysis. These datapoints include 10 values
calculated from spectrographic plates obtained from  Lick Southern Station, Chile, between 1907 and 1911, and  32 RVs calculated from several spectral lines measured with the Radcliffe 74-inch
reflector at Radcliffe Observatory during 1964--1970. We recalculated more accurate HJD values for the Lick data, based on the information given by Campbell \& Moore (1928) and discovered that in their ambiguous notation the RVs of primary and secondary of nights HJD 2417703 and HJD 2418760 were exchanged.
For 25 of the 42 datapoints, RVs for both components are given. 
Also three RV measurements (1955--1956), provided with errors, are
available from Buscombe \& Morris (1960).
A journal of all available RVs we used for the further analysis is given in Table~\ref{RVjournal}, while the individual RVs with corresponding HJDs can be found in Table~\ref{onlinedata}\footnote{Table~\ref{onlinedata} is only available in electronic form.}. The error estimates,
given in Table~\ref{onlinedata} only for our observations, are the rms errors of the mean
of \SiIII, \MgII, and \HeI\ RVs.

\subsection{Determination of the orbital parameters}
Analysis of RVs was carried out using the  \FOTEL code (Hadrava 1990), which is designed to solve the light- and/or RV curves of binary
stars with a possible third component and also can model secular changes in the orbital period and apsidal motion. To each individual RV we assigned a weight which is proportional to the spectral resolution $R$: $w=R/32000$. The spectral resolution can be expressed by the formula
\begin{equation}
R=  {\lambda\over{n\cdot D\cdot s}},
\end{equation}
\noindent with  $\lambda$  the central wavelength of the spectrum in question
in \AA, $D$  its linear dispersion in \AA/mm$^{-1}$, $s$  the ``pixel" spacing in mm,
and $n$  the number of d$\lambda$ per FWHM of the projected slit width, which has a typical value of 2-3. We adopted the values $s$=0.020~mm and $n$=2 for the photographic spectra. Information on the dispersion or resolution of the different types of data is provided in Table~\ref{RVjournal}.
 
First, we calculated a local solution for the subsets of Radcliffe, Lick, and the combined FEROS and GIRAFFE data. In the case of the Lick data we kept the values of the orbital period and eccentricity fixed. For the new data,
measurements with errors larger than 10 km s$^{-1}$ were rejected.
These uncertain measurements correspond to datapoints near
conjunctions. The lack of accurate datapoints close to the system velocity
due to blending of the lines is a typical problem that occurs for
SB2s. One of the consequences is that orbital parameters whose determination is sensitive to the velocity behaviour near conjunction, such as the eccentricity $e$ and the periastron longitude $\omega$, will be less accurately derived.
The resulting parameters of the sub-datasets
are listed in Table~\ref{epsluporbsolution1}.
From the old dataset of Lick and Radcliffe
measurements  we found a solution consistent with the one T70 derived from the Radcliffe data  ($P_{\rm orb}=4.\!\!^{\rm d}5597, e=0.26$). 
The RVs of the FEROS and GIRAFFE time-series could be folded with the same orbital period with a slightly higher eccentricity $e=0.311$.
We note that the values of $\omega$  differ in the different datasets and hence suspected the presence of an apsidal advance. 

Secondly, we searched for solutions in larger subsets, thereby allowing for apsidal advance. We worked iteratively in cases of subsets including datasets 4-6: we determined values of $P_{\rm orb}$ and $\dot{\omega}$, which were subsequently kept fixed in the KOREL disentangling process of all FEROS, GIRAFFE, and CAT spectra (see Sect.~\ref{koreldisentangling}). As KOREL allows  construction of a precise RV curve for both components, even at phases near conjunction, we subsequently ran the FOTEL code again, replacing the FEROS, GIRAFFE, and  CAT measurements by the KOREL RVs to optimise the solution.

We investigated the rate of the rotation of the line of apsides in the subsets including Lick and Radcliffe data (datasets 1 and 3, third column Table~\ref{epsluporbsolution1}) and Radcliffe, CAT, FEROS, and GIRAFFE data (datasets 3-6, using \KOREL RVs for datasets 4-6, second column Table~\ref{epsluporbsolution2}). 
The bad phase coverage of the Lick data, in combination with its poor quality,  did not allow good definition of $\omega$, so we did not find a significant apsidal advance in the first subset.
Next, we combined all datasets, spanning some 96 years, including the few Mount Stromlo and CAT velocities, and searched for the best matching orbital solution. We did not succeed in including the few RV measurements
published by Buscombe \& Morris (1960), as they fell completely
outside our velocity curve. Therefore we question the correctness of
these values. The more so because Buscombe \&
Kennedy (1962)  published an unreliable orbital solution with a
nearly circular orbit of 0.9 days.  Leaving out the Mount Stromlo measurements led to an orbital period of $P_{\rm orb} = 4.\!\!^{\rm d}55983$ and $\dot{\omega} = 0.8 \pm 0.2^{\circ}$/yr, the latter value being slightly smaller than obtained from the datasets 3-6. 
 Including the corresponding \KOREL RVs for CAT, FEROS, and GIRAFFE data led to the 'best' orbital parameters, i.e.\ with the lowest rms value, given in the first column of Table~\ref{epsluporbsolution2}. We stress, however, that although the solutions are quite stable concerning parameters $P_{\rm orb}$, $e$, $K_1$, and $K_2$, uncertainties surround the values of $\omega$, $\dot{\omega}$ and consequently also $T_0$. In order to gain clarity, additional datasets, strategically chosen in time, are needed.

The phase diagrams of Lick \& Radcliffe data (4th column, Table~\ref{epsluporbsolution1}), FEROS, \& GIRAFFE data (last column, Table~\ref{epsluporbsolution1}) and FEROS, GIRAFFE, and CAT data (\KOREL solution, last column, Table~\ref{epsluporbsolution2}) with respect to $P_{\rm orb}$ are displayed in Fig.~\ref{epslupfotelplot}.  

The system velocities of the newly obtained data differ significantly from the older dataset 3 (see Table~\ref{epsluporbsolution2}), which supports the presence of a third body. Given the solution of the close binary system as derived from the
whole available dataset, we attempted
to  converge to a solution of the triple system, but -- not surprisingly, given the poor phase coverage -- without
satisfactory results.  To unravel the triple system with an orbital
period of approximately 64 years, a dedicated long-term project is
required with the aim of  gathering datapoints that are well-spread in the orbital phase.

\begin{figure}
\begin{center} 
\resizebox{\hsize}{!}{\rotatebox{-90}{\includegraphics{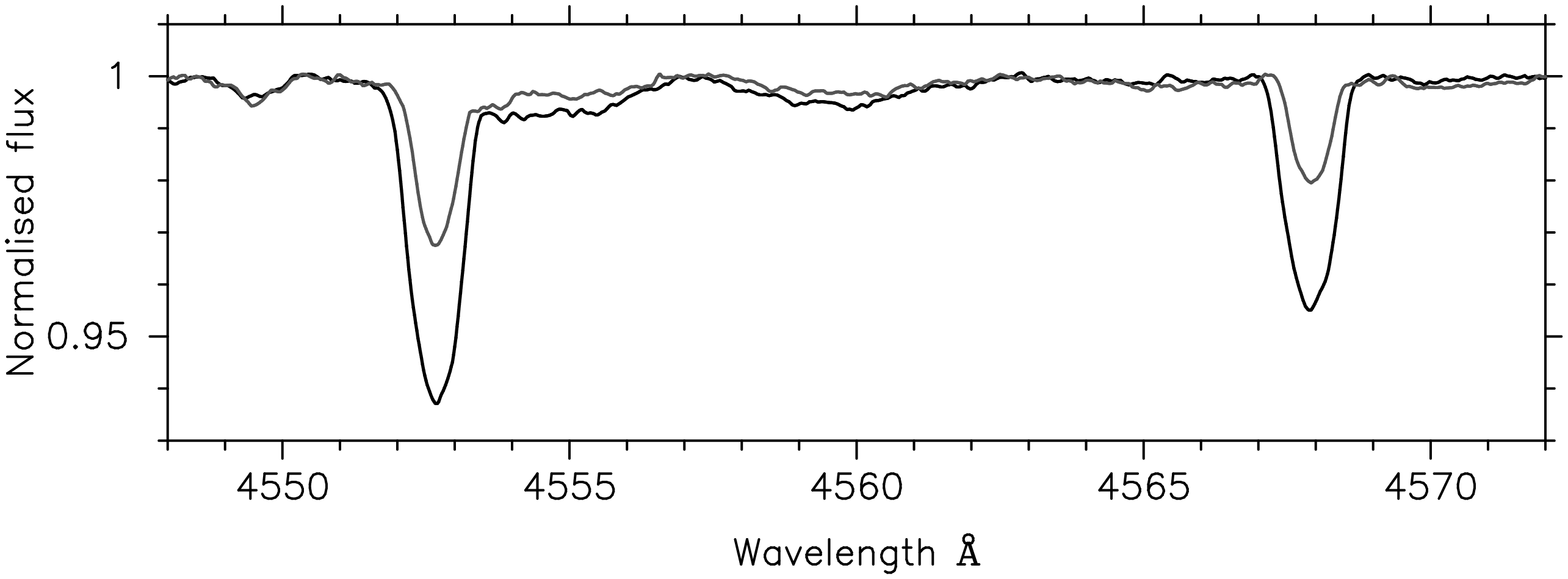}}} 
\caption{Normalised \KOREL disentangled profiles of the \SiIII\ 4553~\AA\ and 4568~\AA\ absorption lines of primary (black) and secondary (gray).}
\label{disentangledepslup}
\end{center}
\end{figure}

\subsection{Spectral disentangling}\label{koreldisentangling}
An advantage of the \KOREL code (Hadrava 1995) is that it allows  the study of a spectral range as a whole and thereby can extract  information from the blended lines. 
We applied the \KOREL disentangling technique to the set of CAT, FEROS, and GIRAFFE spectra and focussed on the wavelength region centered on the two bluest lines of the \SiIII\ triplet. One spectrum was removed from the dataset due to its poor quality.
 We  included
apsidal motion to account for the time gap of 2574  days between the CAT data and the data taken in May 2003. We note, however, that given the small amount of CAT spectra, the solutions allowing for apsidal advance do not differ significantly from the ones not taking apsidal motion into account.
 We kept the orbital period fixed on its best \FOTEL solution ($P_{\rm orb,ano} = 4.\!\!^{\rm d}55983$) and searched for the most satisfactory solution of orbital elements in combination with the 'best' decomposition of the spectra.  The resulting elements are listed in the last column of Table~\ref{epsluporbsolution2}, and the derived RVs of primary and secondary are plotted versus orbital phase in the lower panel of Fig.~\ref{epslupfotelplot}.  
The normalised disentangled spectra centered at 4560~\AA\ are given in
Fig.~\ref{disentangledepslup}. 

\section{Limits on physical elements}

\label{epslupphys}
\subsection{Estimates of basic physical elements}

\begin{figure}
\begin{center}
\resizebox{0.47\textwidth}{!}{\rotatebox{-90}{\includegraphics{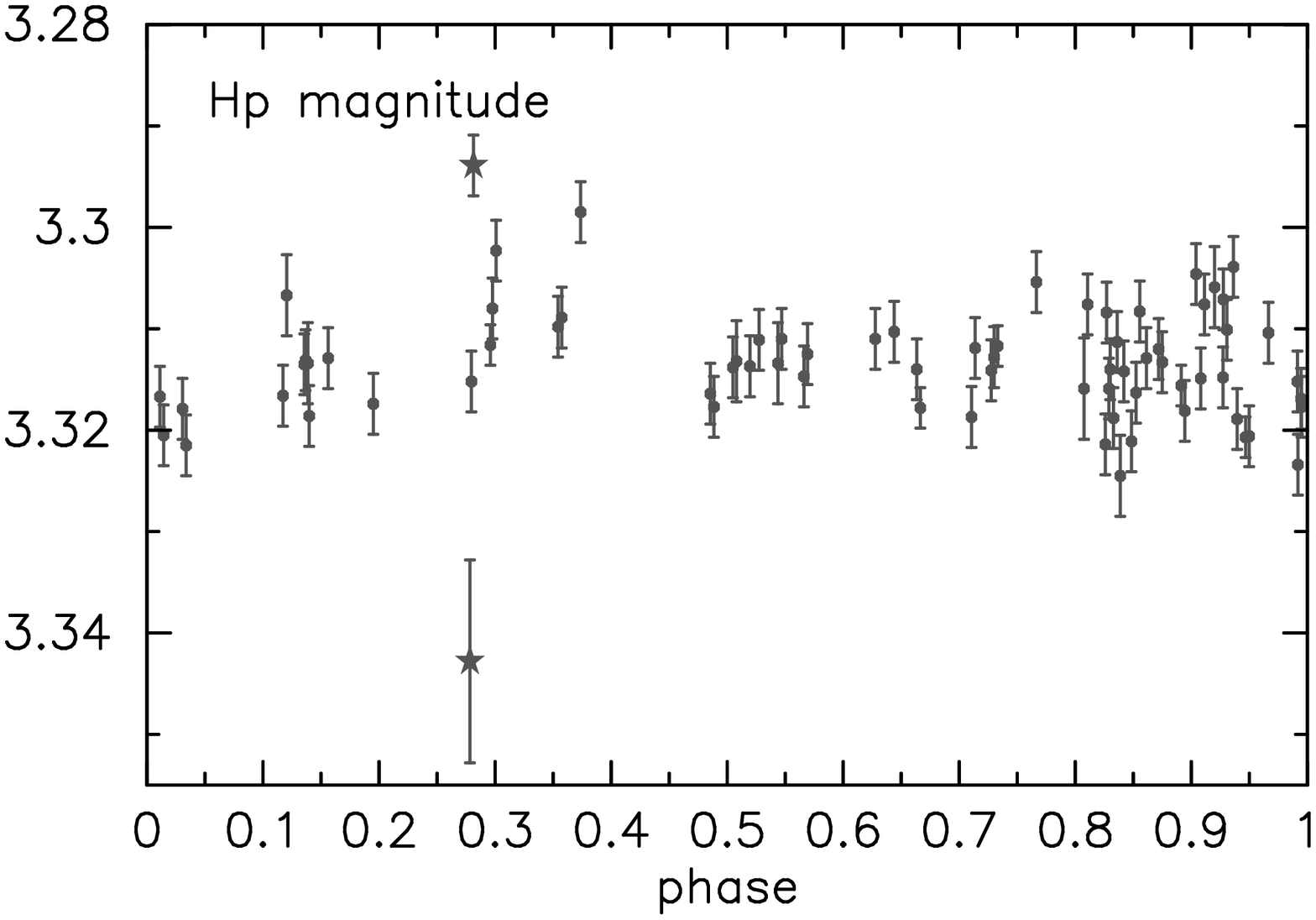}}} 
\caption{The HIPPARCOS data folded according to the close orbit of
$\varepsilon\,$Lup ($P_{\rm orb}~=~4.\!\!^{\rm d}55970$; zero phase corresponds
to the epoch of periastron HJD 2447912). The maximum and minimum
values are indicated by an asterisk.}
\label{hipparcosepslup}
\end{center}
\end{figure}
\begin{figure}
\begin{center}
\resizebox{0.47\textwidth}{!}{\rotatebox{-90}{\includegraphics{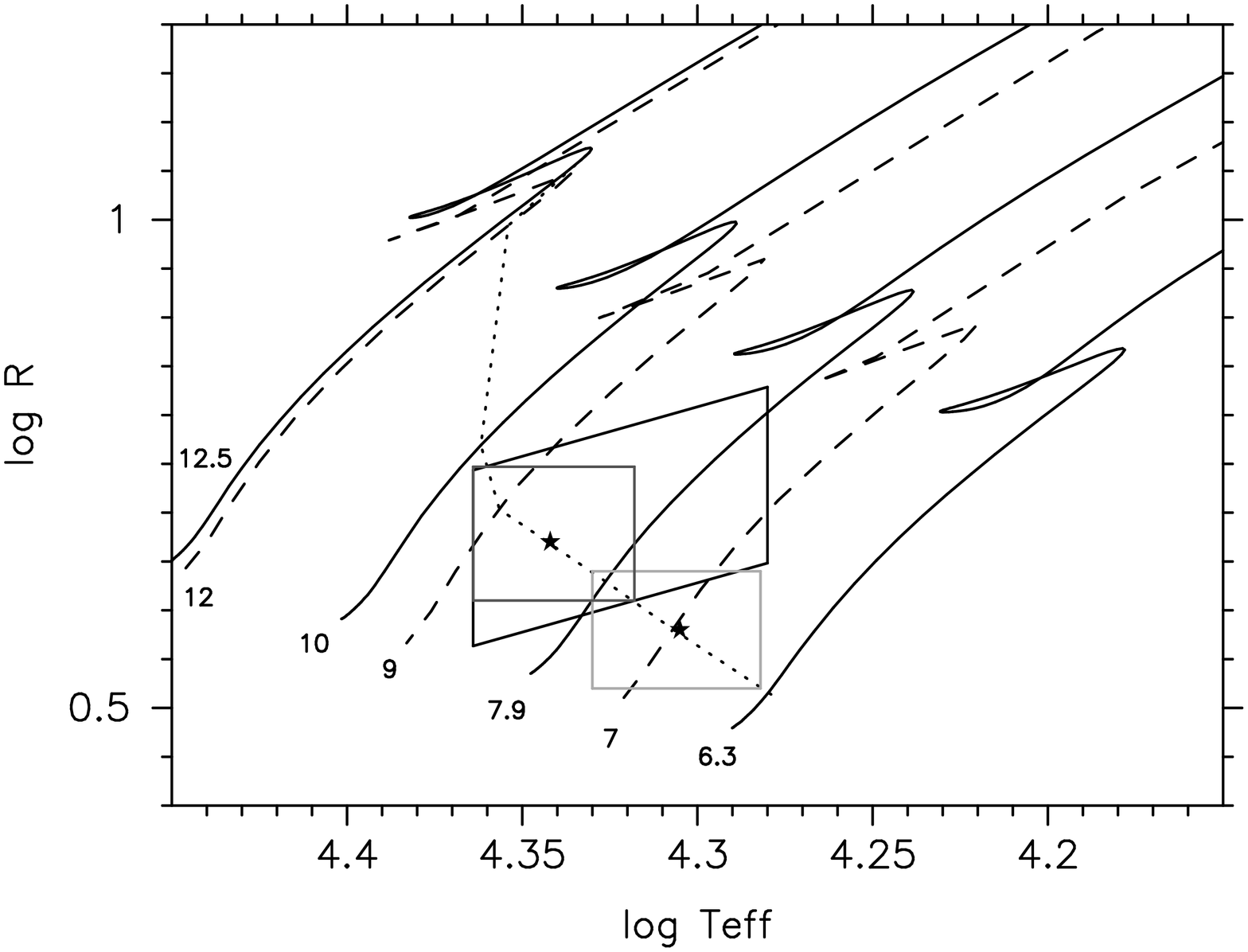}}} 
\caption{$\log T_{\rm eff}$--$\log R/R_{\odot}$ diagram for Schaller's et al.\ (1992, dashed lines) and Claret's (2004, full line) evolutionary models with masses between 6.3$M_{\odot}$ and 12.5$M_{\odot}$, as indicated in the figure. The dotted line represents stars with an age of 15 Myr. The black box  indicates the position of the primary of $\varepsilon\,$Lup assuming $\log T_{{\rm eff},1} \in [4.280;4.364]$ K, and $R_1 \in [3.6,6.7] R_{\odot}$. The dark gray box gives the restricted area assuming an age of approximately 15 Myr, while the light gray box is the corresponding area of the secondary of $\varepsilon\,$Lup. The two asterisks indicate the position of the representatives of the two components of $\varepsilon\,$Lup.}
\label{schallermodel}
\end{center}
\end{figure}

 An important unknown in the unravelling of the physical parameters of binary components is the orbital inclination, which can only be accurately determined in case the star is an eclipsing binary. 
In the literature no evidence for the presence of eclipses in $\varepsilon\,$Lup is found.
We looked at the HIPPARCOS lightcurve of $\varepsilon\,$Lup. The data with
quality label 0 are plotted versus orbital phase in
Fig.~\ref{hipparcosepslup}. We see that one datapoint shows a decrease
in brightness. This happens to be the measurement with by far the largest error bar, which is moreover measured 15 minutes apart from the brighest datapoint in the dataset, spanning more than 1000 days. Therefore we consider
this  insufficient evidence to claim the presence of an
eclipse.  Assuming that $\varepsilon\,$Lup is a non-eclipsing system  implies an orbital inclination lower than 50 degrees ($i_{\rm orb}<50^{\circ}$), given the B-type nature of the two components. 

We estimated the radius $R$ of the primary by using the formula
\begin {equation}
\log (R/R_\odot) = 
7.474-2\log T_{\rm eff}-0.2BC-V+A_{\rm V}-\log\pi,\label{polo}
\end{equation}
with $\pi$ the HIPPARCOS parallax ($6.06 \pm 0.82$ mas, Perryman et al.\ 1997), $BC$ the  bolometric correction, reddening $A_{\rm V}$, effective temperature $T_{\rm eff}$, and $V$ magnitude. From the HIPPARCOS photometry transformed to the Johnson $V$ magnitude by using Harmanec's (1998) transformation formula and from the Geneva photometry, we find $V$ = 3$^{\rm m}\!\!.$375 and $V$ = 3$^{\rm m}\!\!.$369, respectively. The corresponding dereddened value of the latter is $V_0=$3$^{\rm m}\!\!.$32. Next, we considered two  situations: that the two components are equally bright in the $V$ magnitude, implying $V_{0,1}=$4$^{\rm m}\!\!.$06, and that the secondary is half as bright as the primary, hence $V_{0,1}=$3$^{\rm m}\!\!.$75. These two situations are realistic limits on the true value of $V_{0,1}$.  Considering a range of $\log T_{{\rm eff},1} \in [4.280;4.364]$, which is consistent with a spectral type between B3 and B2 according to Harmanec's tabulation (1988), formula~\ref{polo} leads to a radius of the primary between 3.6 and 6.7 $R_{\odot}$ (black box in Fig.~\ref{schallermodel}).

By comparing our observations with Schaller's et al. (1992) and Claret's (2004) evolutionary models (see Fig.~\ref{schallermodel}), and assuming an age of 15 Myr (dotted line) of the $\varepsilon\,$Lup system as a member of the UCL OB association (De Zeeuw et al. 1999), we find the following restrictions on the radius, mass and  $\log T_{\rm eff}$ of the primary of $\varepsilon\,$Lup: $M_1 \in [7.6,9.7] M_{\odot}$, $\log T_{{\rm eff},1} \in [4.32;4.36]$ and $R_1 \in [4.1,5.6] R_{\odot}$ (dark gray box in Fig.~\ref{schallermodel}). Using this range of masses for the primary we can narrow down the interval of the orbital inclination and of the mass of secondary.
 From the values $M_1\sin^3i_{\rm orb}\,\sim\,0.37 M_{\odot}$ and $M_2\sin^3i_{\rm orb}\,\sim\,0.31 M_{\odot}$, derived from  the best \FOTEL
solution, we find $i \in [19.5;21.5]^{\circ}$ and $M_2 \in [6.4;8.2] M_{\odot}$. These values of $M_i \sin^3 i$ are thus compatible with the lack of observational evidence for eclipses. The evolutionary tracks (see Fig.~\ref{schallermodel}) for a secondary with $M_2 \in [6.4;8.2] M_{\odot}$ and the same age as the primary, impose the following limits: $\log T_{{\rm eff},2} \in [4.28;4.33]$ and $R_2 \in [3.5,4.4] R_{\odot}$ (light gray box in Fig.~\ref{schallermodel}). The physical parameters of representatives of the primary and secondary of $\varepsilon\,$Lup, indicated by  asterisks in Fig.~\ref{schallermodel} and  given in Table~\ref{elements}, are compatible with a B3IV and B3V star as suggested by T70. 

\begin{table}
\caption{Representative values of $\log T_{{\rm eff}}$, radius, and mass for the components of $\varepsilon\,$Lup (indicated by  asterisks in Fig.~\ref{schallermodel}), and the associated estimate of the orbital inclination and semi-major axis $a$.}
\label{elements}
\begin{flushleft}
\begin{tabular}{l|lll|ll} 
\noalign{\smallskip}\hline
\hline\noalign{\smallskip}
 star& $\log T_{{\rm eff}}$ & $R/R_{\odot}$ & $M/M_{\odot}$ & $i_{\rm orb}$ ($^{\circ}$) & $a$ ($R_{\odot}$) \\ \hline
 1 & 4.34 & 4.7 &  8.7 & 20.5 & 29.2 \\
 2 & 4.31 & 3.8 &  7.3 &      &       \\ 
\noalign{\smallskip}\hline
\end{tabular}
\end{flushleft}
\end{table}

From the Full Width Half Intensity (FWHI) of the disentangled profiles
(Fig.~\ref{disentangledepslup}), we measured projected rotational
velocities according to the formula $v_1 \sin i_{\rm rot} = {\rm FWHI}/\sqrt{\rm{ln} 16}$, and arrive at values $v_1 \sin i_{\rm rot} = 42$ km s$^{-1}$ and $v_2 \sin
i_{\rm rot} = 32$ km s$^{-1}$ for the primary and secondary, respectively. These values are compatible with the value of 40 km s$^{-1}$ Slettebak et al. (1975) listed in their catalogue.
Assuming $i_{\rm rot} \in [19.5;21.5]^{\circ}$ leads to
the following limits on the equatorial rotational velocities: $v_1 \in [115;125]$ km s$^{-1}$ and $v_2 \in [88;95]$ km s$^{-1}$. 
For a primary and secondary with radii as derived above, we estimate a rotational period between $1.\!\!^{\rm d}7$ and $2.\!\!^{\rm d}7$ days ($f_{\rm rot} \in
 [0.41;0.60]$ c~d$^{-1}$), respectively, $1.\!\!^{\rm d}9$ and $2.\!\!^{\rm d}5$ days ($f_{\rm rot} \in [0.39;0.54]$ c~d$^{-1}$). 
Given all uncertainties involved, it is quite possible that both components
have {\sl identical} rotational periods, possibly
corresponding to the spin-orbit synchronization at periastron,
observed for many eccentric-orbit binaries. For $\varepsilon\,$Lup
its value is $2.\!\!^{\rm d}$481 if the eccentricity $e=0.277$
from the KOREL solution is adopted (cf. formula 7 and relevant
discussion in Harmanec 1988).

We have to bear in mind that, due to several limitations on our dataset, the derived values of $M_i$, $R_i$, $T_{{\rm eff},i}$ ($i=1,2$) and the orbital inclination are only rough estimates, derived under certain assumptions.  Additional information from future interferometric observations would provide  a much more accurate determination of the component masses and radii, as well as the orbital inclination and the relative luminosities of the components. 

\subsection{Interpretation of the observed apsidal motion}
The range of apsidal periods we derived in Sect.~\ref{apsmotion} ($U = 428 \pm 79$) is consistent with the values obtained for other binaries with orbital periods between 4 and 5 days
(Petrova \& Orlov 1999). As a comparison, we mention the  system QX\,Car,
which is very similar to $\varepsilon\,$Lup, and consists of two B2V
stars orbiting in an eccentric ($e = 0.28$) orbit of 4.5 days, which
has an apsidal period of  370 years. 

The
tide-generating potential of the binary system can give rise to a periodic variation of the longitude of periastron. It is well-known that the observed apsidal advance in eccentric-orbit
binaries arises from both  classical and general-relativistic
terms, i.e. \\
$\dot{\omega}_{\rm obs}=\dot{\omega}_{\rm clas}+\dot{\omega}_{\rm rel}$. 
We calculated the contribution of the relativistic term from the formula
given by Gim\'enez (1985),
which resulted in $\dot{\omega}_{\rm rel} \in [0.113;0.139]^{\circ}/$yr and which is a 6-7 times smaller value than our observed range of values.

An estimate of the total contribution of classical effects can be derived from $\dot{\omega}_{clas} = \dot{\omega}_{\rm obs} - \dot{\omega}_{\rm rel}$ and leads to $\dot{\omega}_{clas} \in [0.687;0.661]^{\circ}/$yr. 
The classical effects are caused by tidal and rotational distortions, and, if relevant,  the presence of a third body: $\dot{\omega}_{\rm clas} = \dot{\omega}_{\rm tidal} + \dot{\omega}_{\rm rotation} + \dot{\omega}_3$. 
 The latter term is negligible in the case of $\varepsilon\,$Lup due to
the long orbital period of the triple system. According to the formula
given by Wolf et al.\ (1999), the contribution of the third component
to the rotation of the line of apses is  estimated as $\dot{\omega}_3 = 3 \times 10^{-6}$$^{\circ}/{\rm yr}$. Neglecting this effect, we estimated a mean apsidal constant $k_2$ (cf.\ formula 4 in Claret \& Willems 2002) for the primary and secondary from the observed apsidal period and relative photometric radii
of the components, $r_1=0.16$ and $r_2=0.13$, as $k_2= 0.0074$ for the representative set of values from Table~\ref{elements}. This can be compared with the theoretically expected values, interpolated
from Claret's (2004) models, $k_2=0.0068$ and $k_2=0.0074$ for the primary and secondary, respectively.  It is seen that there is no obvious conflict between the observed and theoretically predicted values. 
Given our observed range of primary radii, assuming $i_{\rm orb}=20.5^{\circ}$, $e=0.277$ and $k_2=0.007$, we estimate the rate of secular apsidal motion due to the tidal distortion of the components from the formula given by Claret \& Willems (2002, their formula 13): $0.004^{\circ}/$yr~$<\dot{\omega}_{\rm tidal}\,<0.019^{\circ}/$yr.
As we could derive only rough estimates of the basic physical properties of both stars, we do not take into account the possible effect
of dynamic tides studied by Claret \& Willems (2002). 
We note, however, that their effect will probably be small as the
stars are rotating close to spin-orbit synchronization at periastron
unless there is some resonance present in the system. This must be left
unanswered until much more accurate masses and radii of components are
known.
The estimated values of $\dot{\omega}_{\rm tidal}$ and $\dot{\omega}_3$ lead us to conclude that the term $\dot{\omega}_{\rm rotation}$, accounting for rotational effects, must contribute significantly to the apsidal motion.

\section{Intrinsic variability}
\begin{figure}
\begin{center}
\begin{tabular}{c}
\resizebox{0.41\textwidth}{!}{\rotatebox{-90}{\includegraphics{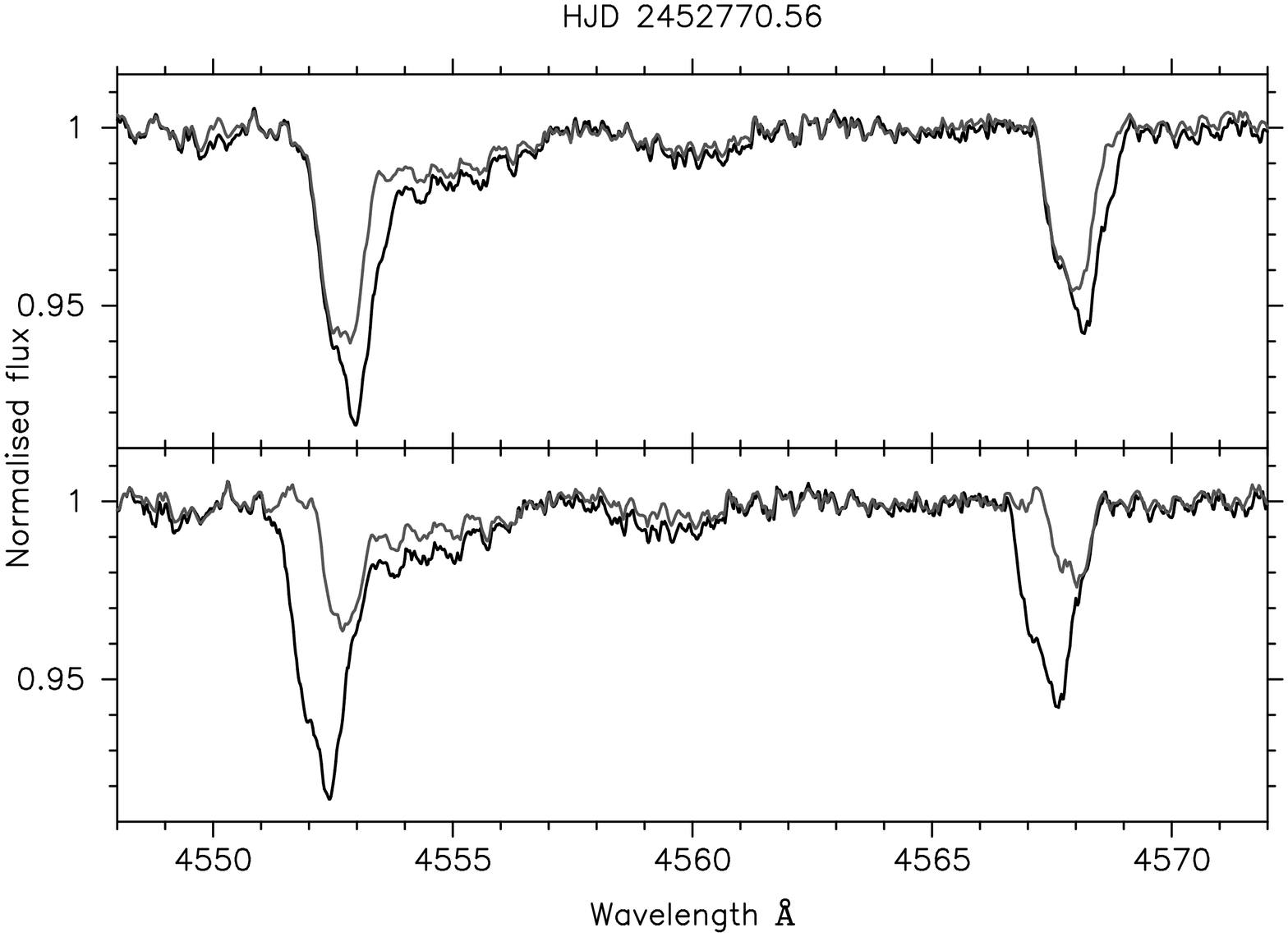}}} \\
\resizebox{0.41\textwidth}{!}{\rotatebox{-90}{\includegraphics{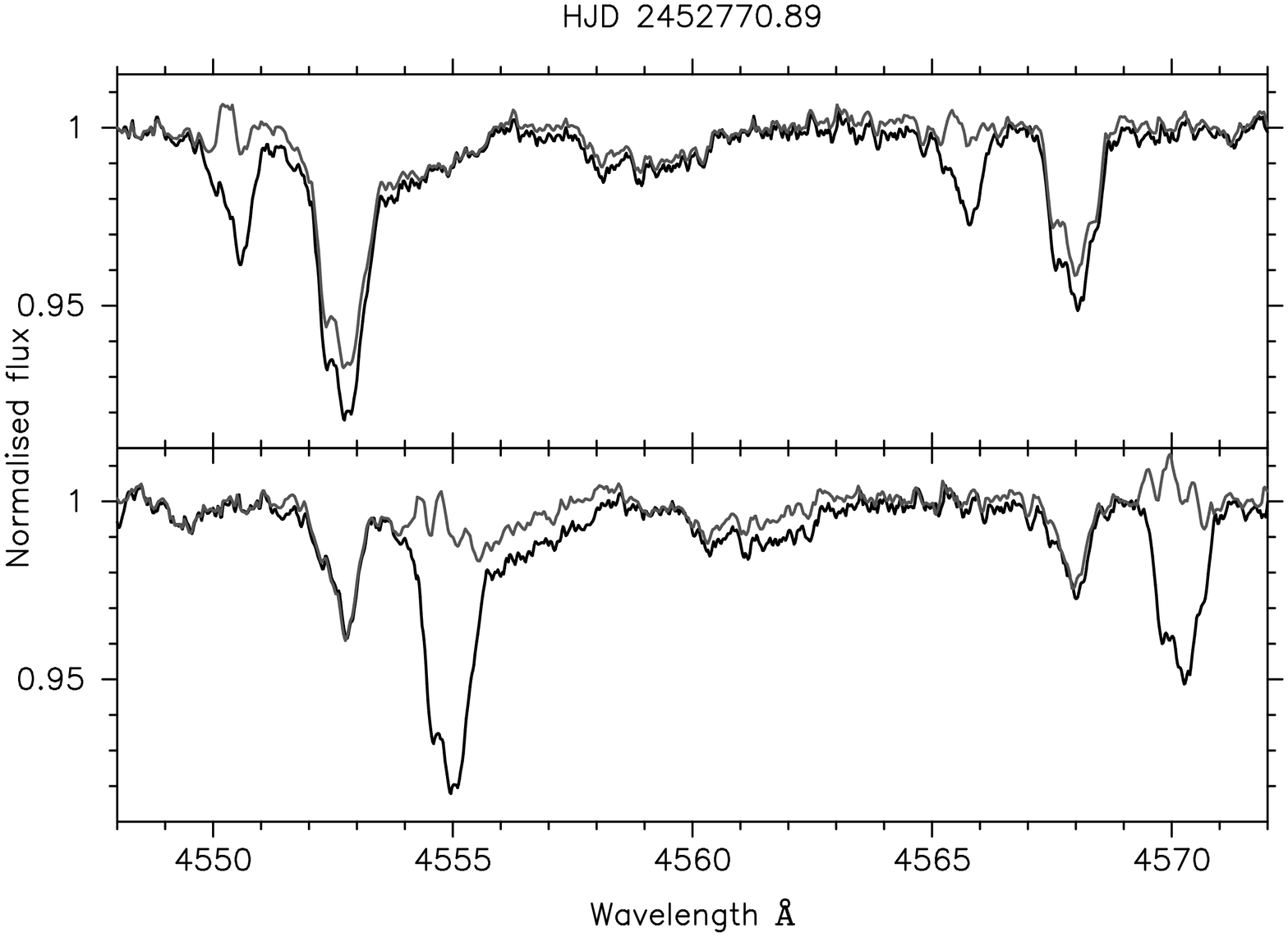}}} \\
\resizebox{0.41\textwidth}{!}{\rotatebox{-90}{\includegraphics{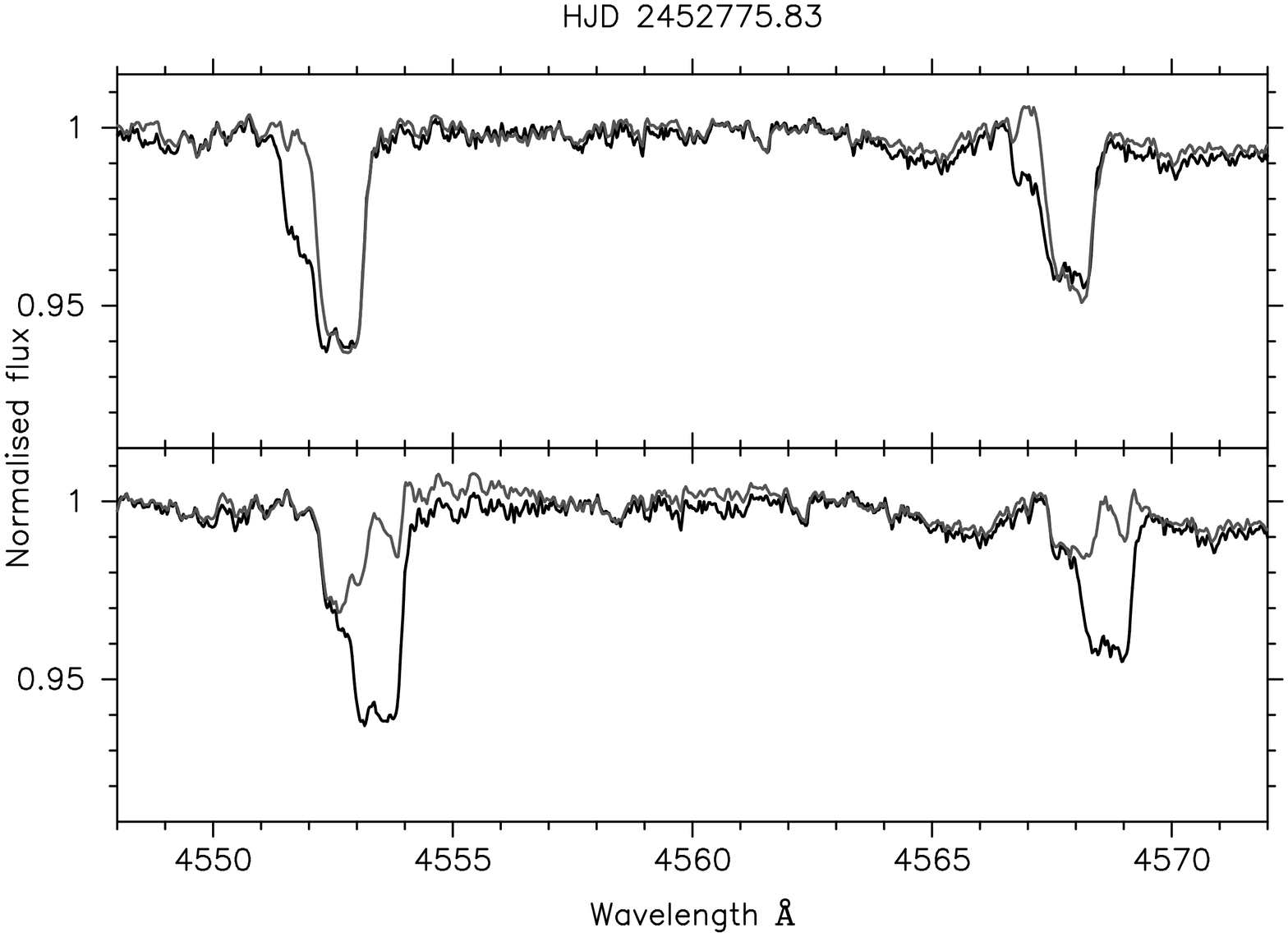}}}
\end{tabular}
\caption{\SiIII\ line-profiles observed at three different phases of the orbital period. The top (bottom) of each panel shows the 'original' profile of the primary (secondary) in black and the 'disentangled' primary (secondary) profile in gray.}
\label{profepslup}
\end{center}
\end{figure}

\begin{figure*}
\begin{center}
\begin{tabular}{cc}
HJD 2452771 &  HJD 2452773 \\
\resizebox{0.4\textwidth}{!}{\includegraphics{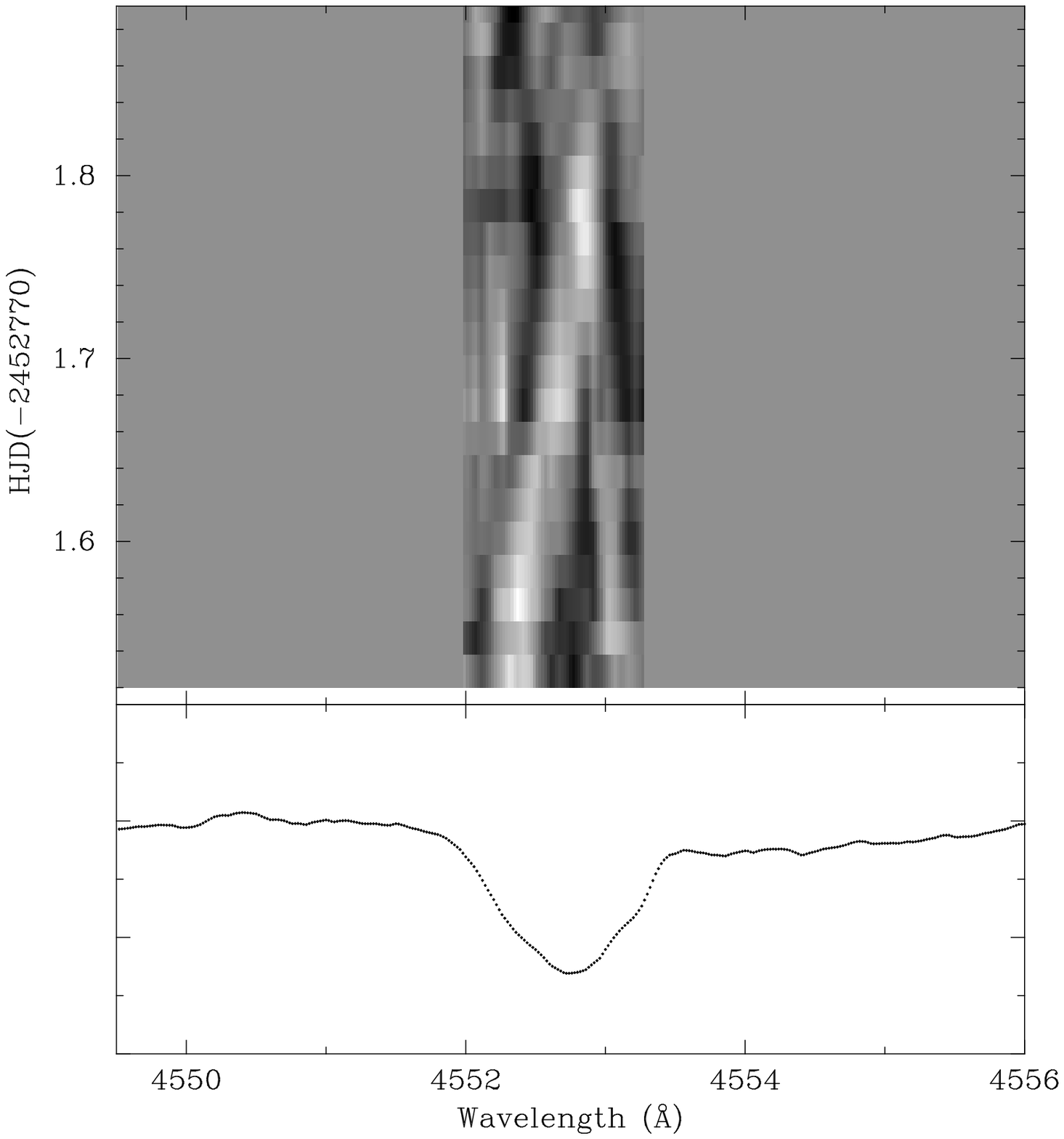}}
&
\resizebox{0.4\textwidth}{!}{\includegraphics{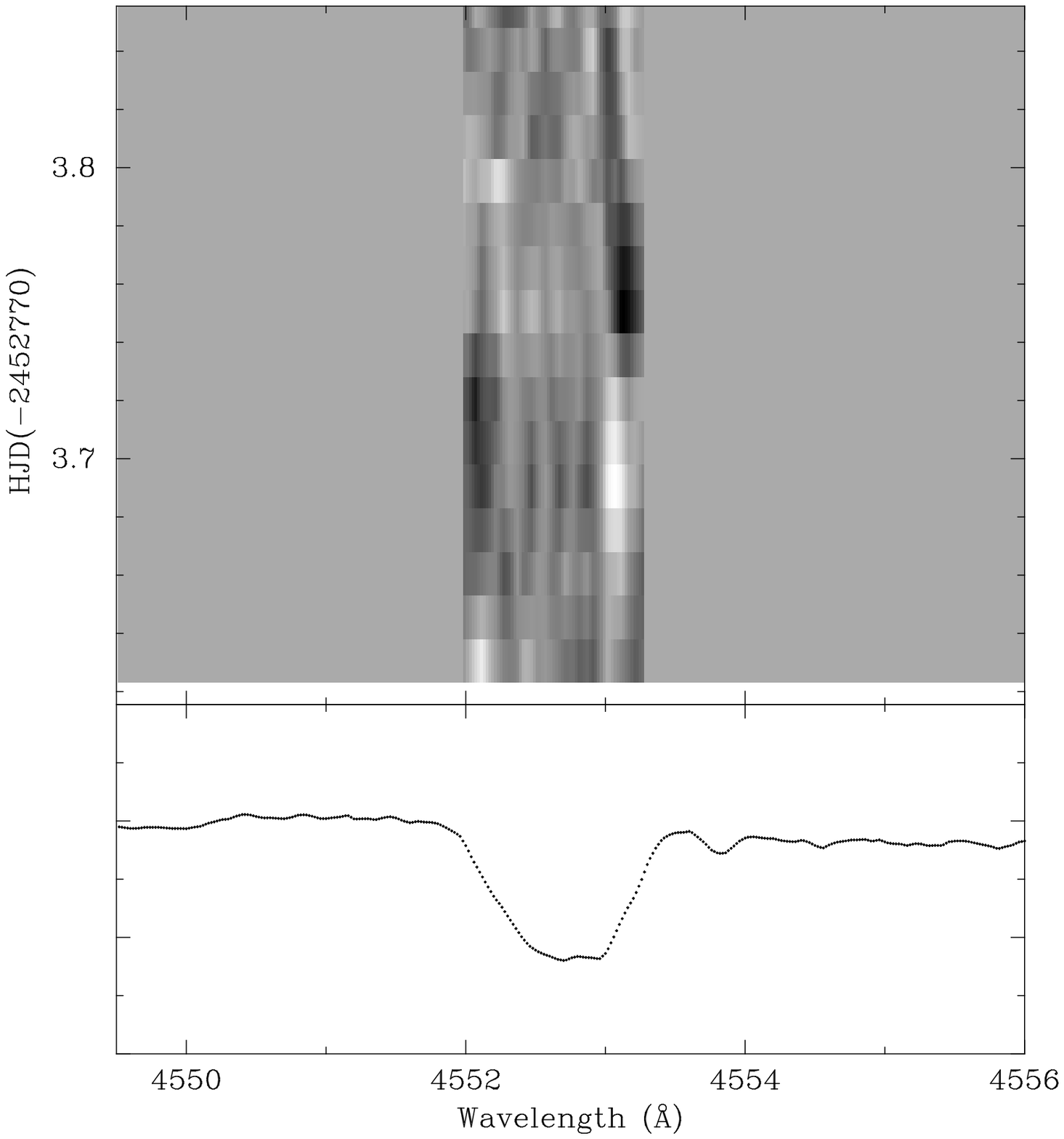}} \\ \\
\resizebox{0.4\textwidth}{!}{\includegraphics{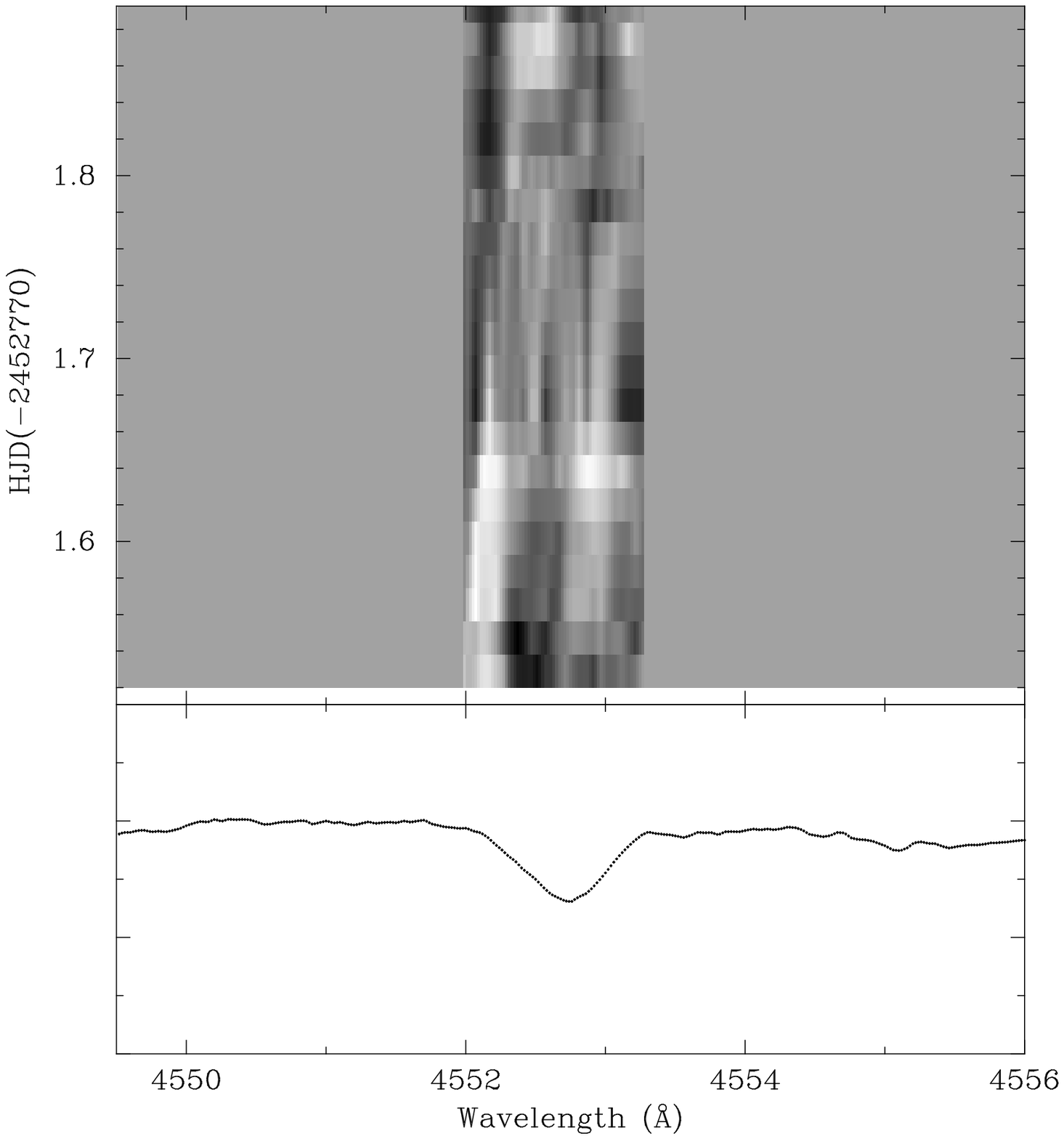}}
&
\resizebox{0.4\textwidth}{!}{\includegraphics{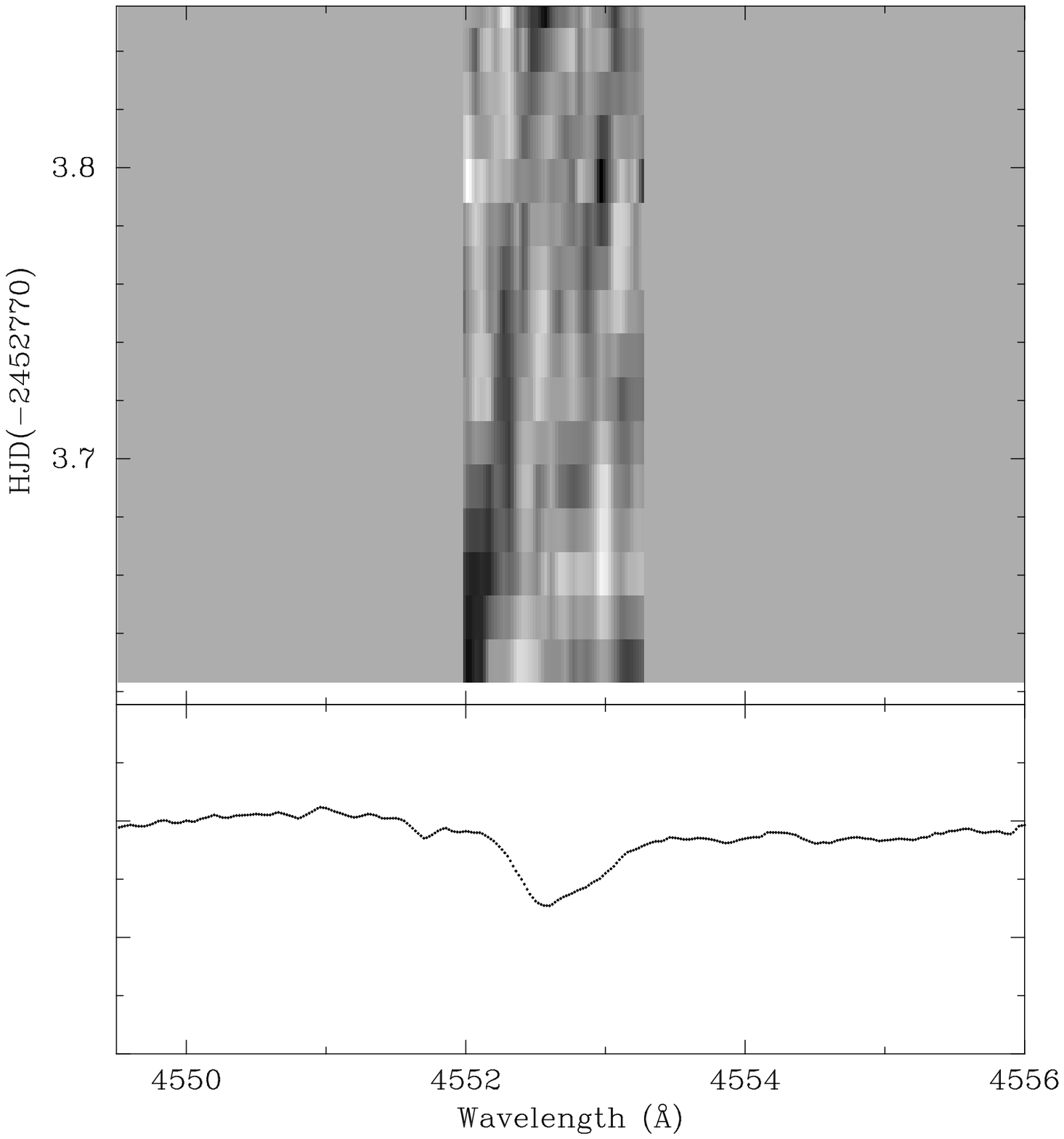}} \\
\end{tabular}
\caption{Grayscale representations of the disentangled FEROS spectra of the primary (top) and secondary (bottom) obtained
during the nights of HJD\,~2452771 (left)
and HJD\,~2452773 (right).   At the bottom of each figure the corresponding nightly average profile is plotted as a reference.}
\label{epslupgraycomp1+2}
\end{center}
\end{figure*}

High-degree ($\ell > 2$) modes are suspected in the primary of $\varepsilon\,$Lup (Schrijvers et
al.\ 2002; Fig.~\ref{CATspectra}).
 The analysis of LPVs of both primary
and secondary is a challenging task due to the comparable
line-strength of the two components and the blending of the component
lines in phases far from elongation. The spectra of two line-profile variable components that have been disentangled by \KOREL have never been the subject of an LPV study before. We investigate the two bluest \SiIII\ profiles before and after \KOREL
disentangling and then compare the results.   A disadvantage of using the original
spectra is that only spectra obtained at phases of
elongation can be used, while all \KOREL disentangled profiles can be
considered for analysis. However we question whether the intrinsic
variations of primary and secondary are properly disentangled, as \KOREL is not constructed to treat  pulsational variations. For the testcase $\kappa$ Sco, Harmanec et al. (2004) and Uytterhoeven et al. (2005) obtained  good results concerning the retrieval of intrinsic variations with high amplitude after disentangling but the performance of \KOREL was less clear for low amplitude variations.

The set of disentangled profiles of the first (second) component were
constructed by adding the \KOREL residuals calculated in the restframe
of the primary (secondary) to the normalised disentangled profile of
the primary (secondary). 
The 'original profiles' of the first (second) component were obtained from the observed spectra by a shift in wavelength according to its corresponding orbital velocity  in the best-fitting orbit obtained with \KOREL (last column in Table~\ref{epsluporbsolution2}, bottom panel in Fig.~\ref{epslupfotelplot}). 
In Fig.~\ref{profepslup} we compare the original and disentangled
profiles of the primary and the secondary at three different orbital
phases. From these comparison plots we  conclude the following: 1. The primary \emph{and} secondary of $\varepsilon\,$Lup are line-profile variables. 2. The \KOREL residuals in the restframe of the primary (secondary) contain signatures of the intrinsic variability of \emph{both} components.  

In order look closely for signs of intrinsic variability, we show the grayscale representations of the disentangled \SiIII~4553 \AA\ profiles of the primary and secondary on two nights during the FEROS
observing run when the two components are near elongation (Fig.~\ref{epslupgraycomp1+2}). To bring
out the variability of the primary and secondary in full detail, we calculated the residuals with
respect to the nightly average spectrum.
For both components we find
signatures of intrinsic
variability, although weak in case of the secondary,  in the form of  moving black and white bands.

\begin{figure}
\begin{center}
\begin{tabular}{c}
\resizebox{0.4\textwidth}{!}{\rotatebox{-90}{\includegraphics{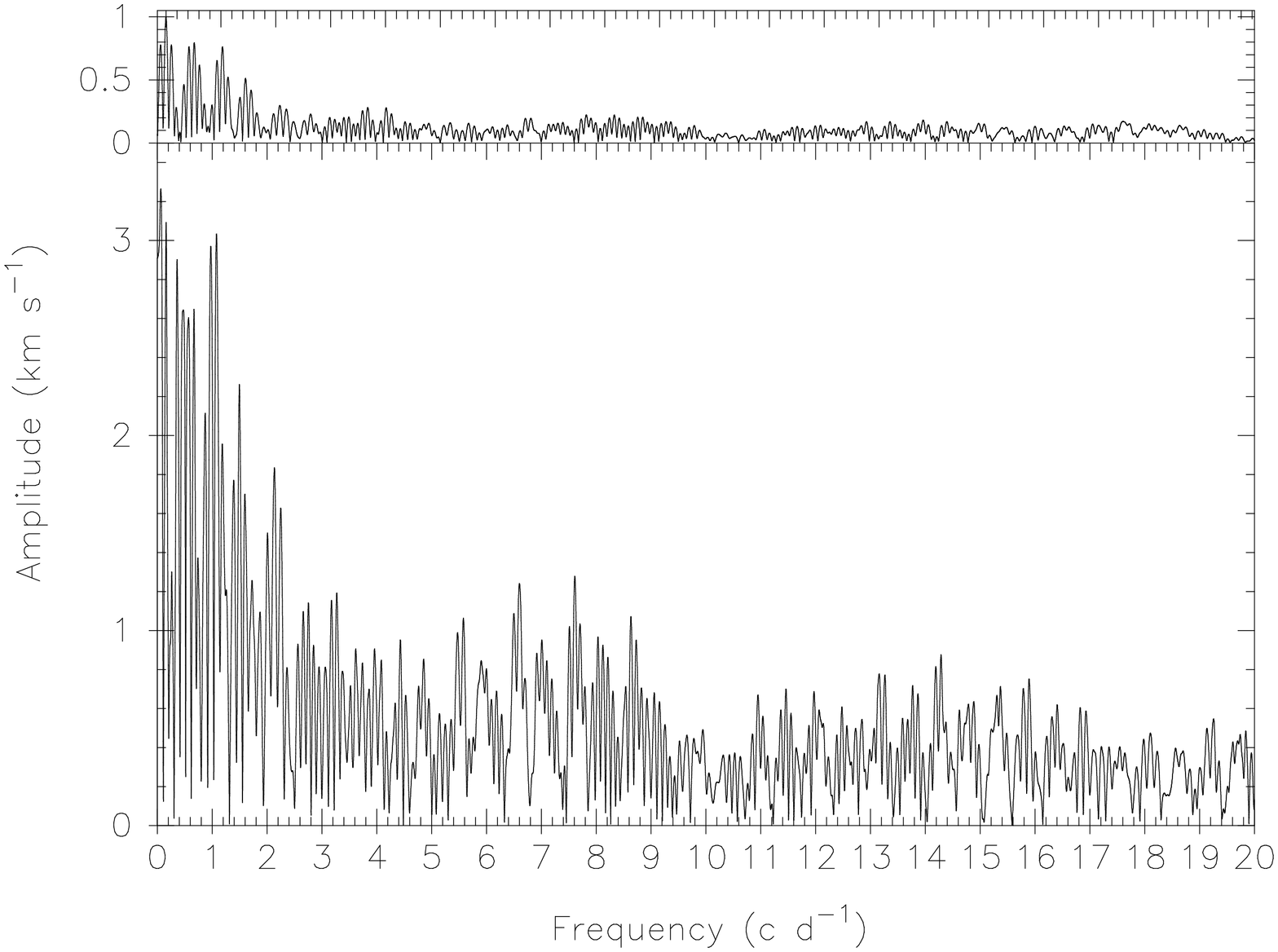}}} \\
\resizebox{0.4\textwidth}{!}{\rotatebox{-90}{\includegraphics{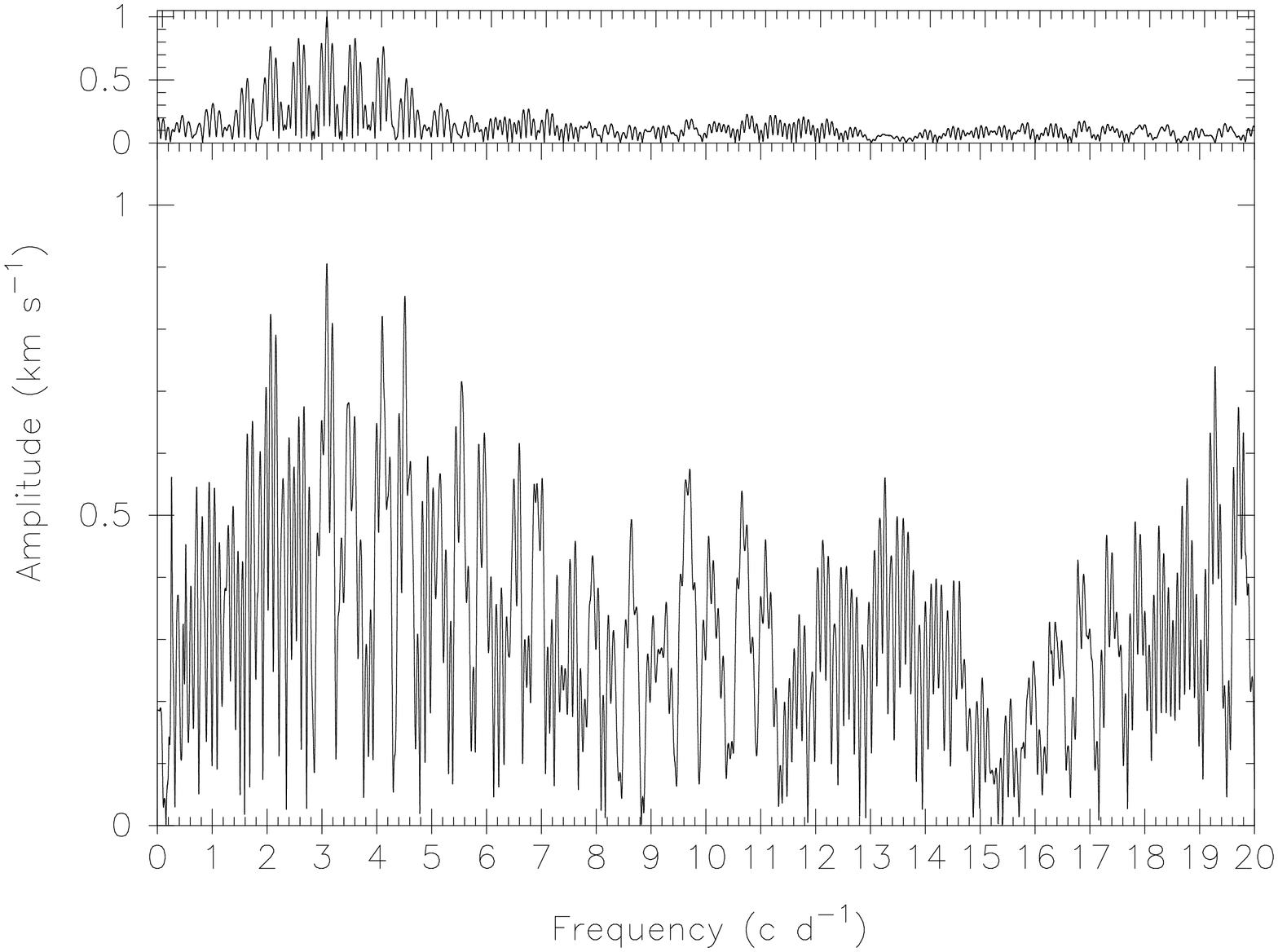}}}
\end{tabular}
\caption{Lomb-Scargle periodograms of $\langle v\rangle$ of
the disentangled  \SiIII~4552.654~\AA\
profiles of the primary. The bottom panel gives the periodogram of
the data after prewhitening with the highest peak at 0.06 c
d$^{-1}$.  The top of each panel gives the window function, shifted towards the position of the
highest frequency peak.}
\label{epslup_sca_comp1b}
\end{center}
\end{figure}

We performed \CLEAN (Roberts et al.\ 1987),
Lomb-Scargle (Scargle 1982) and Phase Dispersion Minimisation (PDM, Stellingwerf 1978) analyses on the first three velocity moments ($\langle v\rangle, \langle v^2\rangle, \langle v^3\rangle$), the EW, and the RVs associated to the minimum of each line-profile, in search for intrinsic variations of primary and secondary.
Also the 2D-analysis Intensity Period Search (IPS, Telting \& Schrijvers 1997), to find variable signals in individual wavelength bins ($\Delta \lambda = 0.02$~\AA\ in case of $\varepsilon\,$Lup), was performed by using the  CLEAN algorithm. 
 All the diagnostics were calculated from the \SiIII~4552.654~\AA\ and 4567.872~\AA\ profiles of both the disentangled and original profiles. 

We investigated the variable signal of the complete datasets,
including all datapoints, and of  reduced datasets, which  included
only the 56 da\-ta\-points obtained when the lines of the component spectra were separate (for a logbook, see Table~\ref{logepslup}). The
clearest signals were detected in the variability of the latter
datasets for the
obvious reason that unreliable datapoints near conjunction were not
included. In theory, all datapoints after
disentangling should contain reliable information about the
variability of the primary (secondary), but  the observations show
that even for the disentangled profiles, this restriction to the 56
separate  spectra is to be preferred. For the moment it is unclear how
well the \KOREL disentangling procedure can separate the intrinsic
variability contributions of the primary and secondary near
conjunction. This can only be deduced from a dedicated simulation
study, which is beyond the scope of the current work.

Below, we report on the results obtained from the reduced datasets
containing 56 datapoints and spanning 12 days. The window
function is very complicated and leads to severe aliasing. The half width at half maximum of the amplitude of the highest window
peak is as large as 0.03 c~d$^{-1}$  and gives  a rough estimate of
the error on the frequencies. According to the empirically derived
formula by Cuypers (1987), the small dataset  allows a frequency  accuracy of 0.01 c~d$^{-1}$. Hence, we searched for frequencies in the frequency domain between 0 and 20 c~d$^{-1}$, with frequency steps of  0.01 c~d$^{-1}$. 
A more extended discussion of the frequency analysis results than presented here can be found in Uytterhoeven (2004)\footnote{PhD thesis available from \\
http://www.ster.kuleuven.be/pub/uytterhoeven\_phd/}.

\subsection{Intrinsic variability of the primary}
No clear dominant frequency is found in the variability of the
primary.   The highest amplitude is found  between 0-2 c~d$^{-1}$.  In the Scargle periodogram of the non-prewhitened
data, as shown in the upper panel of Fig.~\ref{epslup_sca_comp1b}, one
sees hints  of the presence of a variability near
6.46 c~d$^{-1}$ or one of its aliases, but these peaks reduced 
in amplitude (original profiles) or almost completely vanished
(disentangled profiles) after prewhitening with the highest amplitude
low frequency peak
 (lower panel of Fig.~\ref{epslup_sca_comp1b}). Signatures of the peak 6.46 c
d$^{-1}$ are also found at very low amplitude in the IPS analysis. 

The EW of the primary's original profiles  shows variations with
frequency 0.83 c~d$^{-1}$. This value is also recovered in the second moment
and in the IPS analysis, and can be related to 0.44 c~d$^{-1}$, detected in the EW variations of the disentangled
spectra, through a peak of the window function.
As stellar rotational
frequencies are not uncommonly  detected in the EW of line profiles
(e.g.\ $\kappa$ Sco, Uytterhoeven et al.\ 2005; V2052
Oph, Neiner et al.\ 2003; $\beta$~Cephei\,, Telting et al.\ 1997), we
might consider this frequency to be related to the rotation of the
star. Indeed, the value 0.44 c~d$^{-1}$ is compatible with the rotational period assuming spin-orbit synchronization, as derived in Sect.~\ref{epslupphys}.

\subsection{Intrinsic variability of the secondary}

\begin{figure}
\begin{center}
\begin{tabular}{c}
\resizebox{0.4\textwidth}{!}{\rotatebox{-90}{\includegraphics{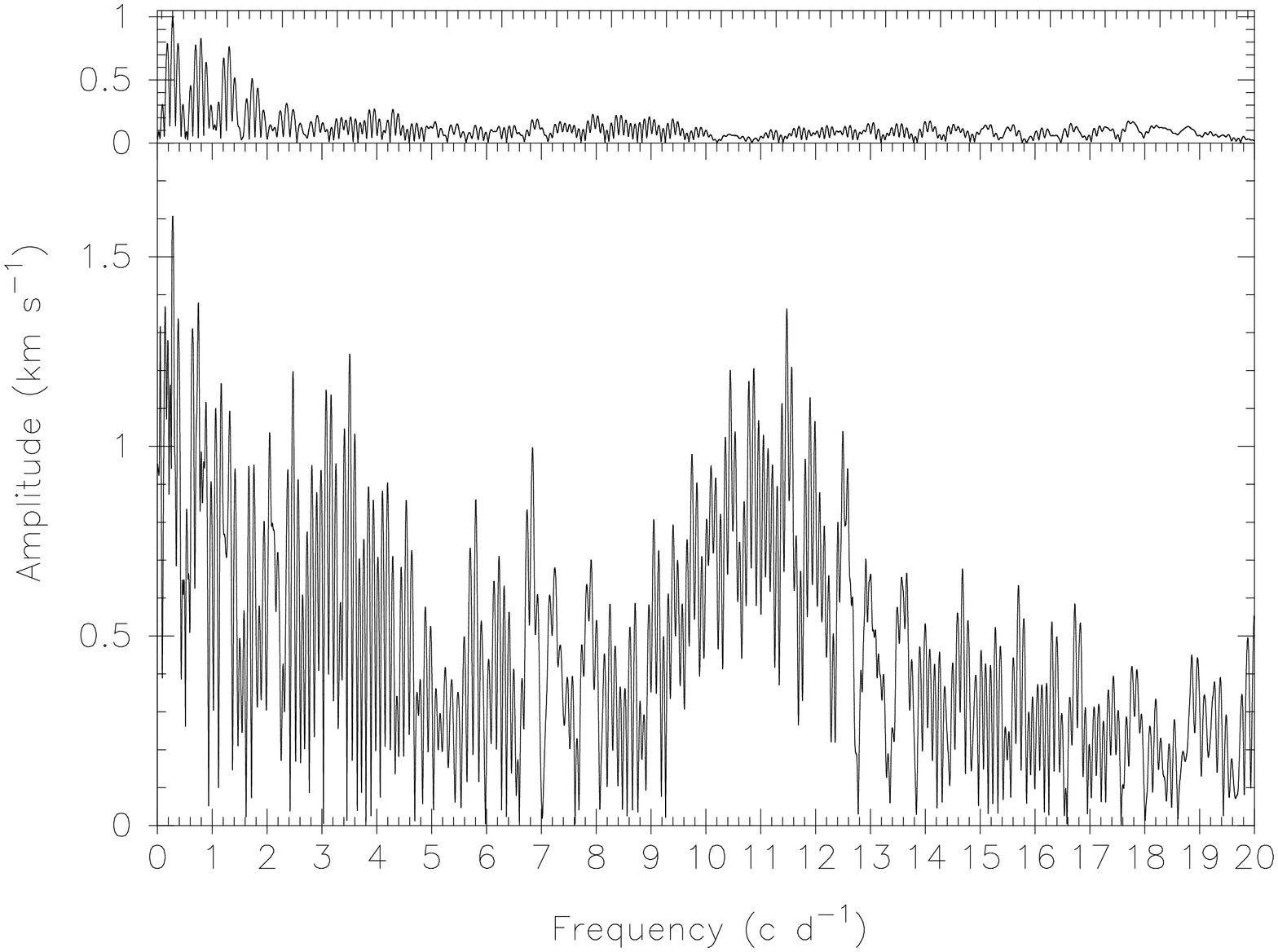}}} \\
\resizebox{0.4\textwidth}{!}{\rotatebox{-90}{\includegraphics{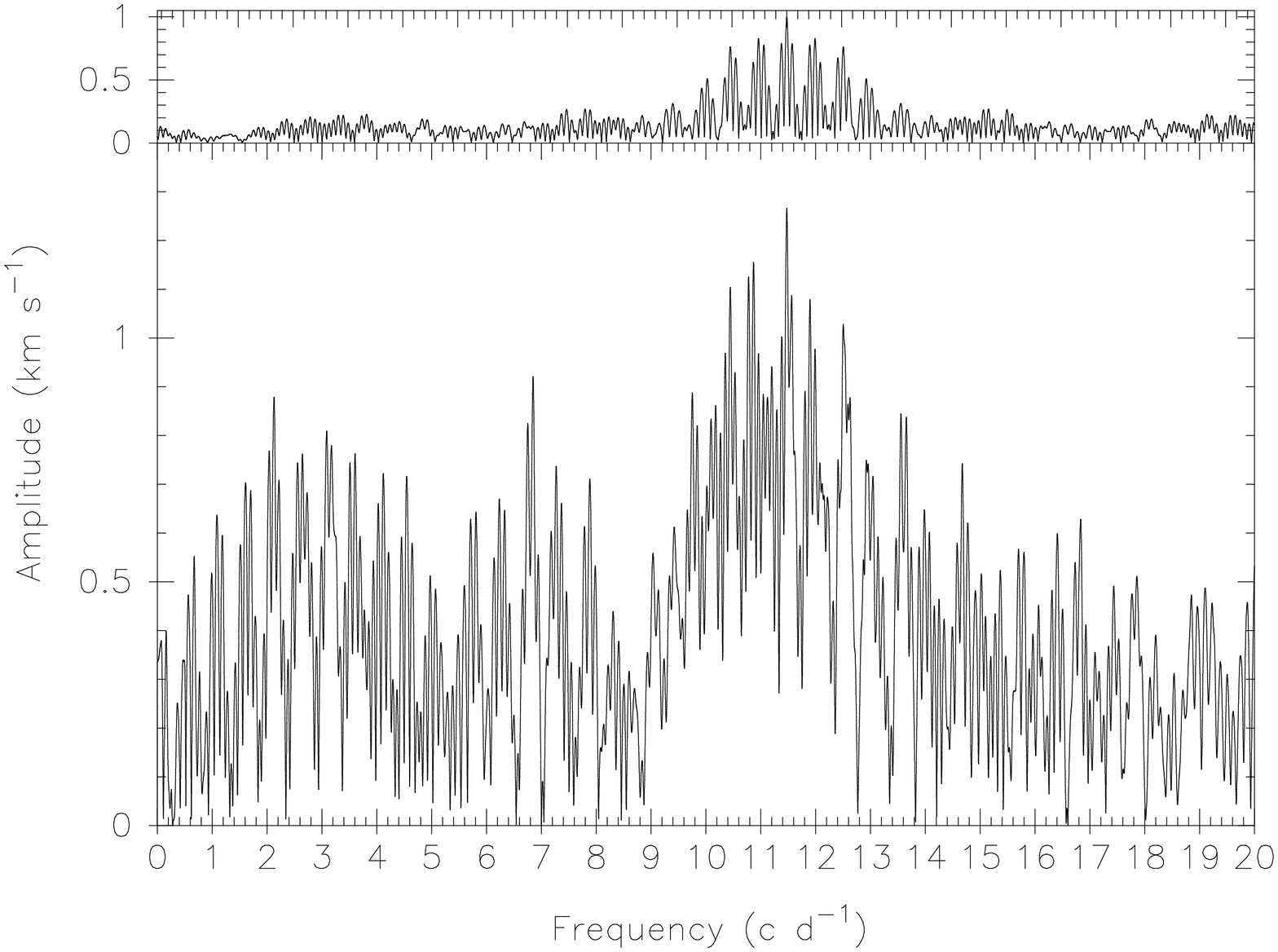}}}
\\
\end{tabular}
\caption{Lomb-Scargle periodograms of the first normalised moment of
the disentangled  \SiIII~4552.654~\AA\
profiles of the secondary. The moments were calculated using fixed
integration boundaries.   The bottom panel gives the periodogram of
the data after prewhitening with the highest peak at 0.28 c~d$^{-1}$.}
\label{epslup_sca_comp2b}
\end{center}
\end{figure}

\begin{figure}
\resizebox{0.85\hsize}{!}{\rotatebox{-90}{\includegraphics{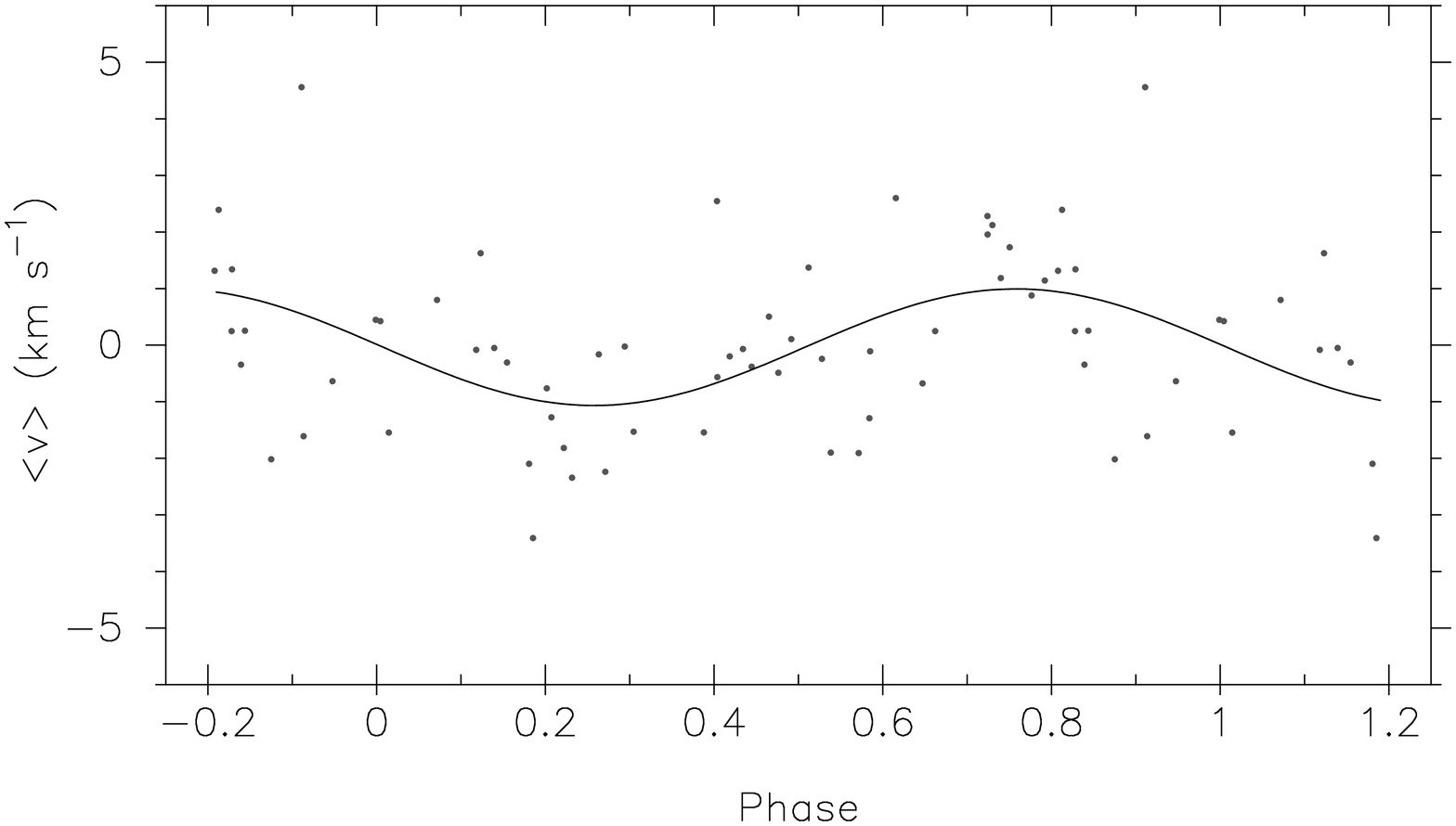}}} 
\caption{The first normalised moment (dots) calculated from the
\SiIII~4552.654~\AA\ profiles of the disentangled spectra of the secondary, after prewhitening with peaks in the interval 0 - 2 c
d$^{-1}$, folded with 10.36 c d$^{-1}$. The full line is the fit with 10.36 c d$^{-1}$.}
\label{phasecomp2}
\end{figure}

The line profiles of the secondary of $\varepsilon\,$Lup\ show
variations with a frequency near 10.36\,c\,d$^{-1}$ (hereafter called $f_1$) and with a velocity amplitude of $1.0 \pm 0.2$ km s$^{-1}$. 
This frequency, or an alias, appears in all diagnostics after prewhitening 
with 0.39 c~d$^{-1}$ or one of its aliases (Fig.~\ref{epslup_sca_comp2b}).
A phase diagram of the first velocity moment of the disentangled
profiles folded with 10.36 c~d$^{-1}$,
is given in Fig.~\ref{phasecomp2}. The frequency 10.36 c~d$^{-1}$ is not recovered
from the IPS analysis.
The EW of the secondary's original profiles varies with 0.75 c~d$^{-1}$, or one
of its aliases. 

We note that the frequency 0.05 c~d$^{-1}$, detected in the IPS
analysis of the profiles of the secondary, also
appears in the IPS analysis performed on the profiles of the primary. This frequency is most probably a reflection
of the orbital motion in the profiles, given the following relation
within the frequency resolution: $0.05 \sim 1/4 f_{\rm orb}$.
Similar relations between intrinsic frequencies and the orbital frequency  are observed in other close binary stars (e.g.\ $\alpha$ Vir, $f = 1/12 f_{\rm orb}$, Smith 1985; $\sigma$ Sco, $f = 1/4 f_{\rm orb}$, Chapellier \& Valtier 1992).

 The  period $1/f_1$
of about 2 hours is relatively short for $\beta$~Cephei\, type
pulsations. A few examples of $\beta$
Cephei stars with similar frequency values are
known (for an overview see Aerts \& De Cat 2003). So far, the highest
frequency  reported in the literature
amounts to 15 c d$^{-1}$ and corresponds to a very low RV amplitude ($\omega^1$ Sco, Telting \& Schrijvers 1998). The low-amplitude
frequency we found in the secondary of $\varepsilon\,$Lup resembles
the low-amplitude frequencies $f'_1 = 10.48$ c d$^{-1}$ and $f'_2 =
10.73$ c d$^{-1}$ detected in the B-type primary of the spectroscopic
binary $\psi^2$ Orionis (Telting et al.\ 2001), which were interpreted
in terms of high-degree pulsation modes.
 To confirm the presence of $f_1$ and refine its value, additional monitoring of $\varepsilon\,$Lup is required. As the intrinsic
variability of  the secondary can only be studied at orbital phases
near elongation, one should aim to gather data during these
phases. 

Frequency $f_1$ shows up in both original and disentangled
datasets. This result indicates that the  variability inherent to the
secondary is well preserved through \KOREL dis\-en\-tangling during 
orbital phases near elongation.

\section{Summary and discussion}

From a dataset of 106 high-resolution spectra obtained
quasi-consecutively from two dif\-fe\-rent observatories with a time-span
of 20 days,  we were able to refine the orbital parameters of the close
orbit of the triple system $\varepsilon\,$Lup by using the \FOTEL and \KOREL
codes. As the star is an SB2, we obtained RVs for both components and found an eccentric ($e=0.277$) orbit of $4.\!\!^{\rm d}55970$ days.  By adding our data to  published RV measurements we found strong evidence of the presence of apsidal motion ($U \sim 430$ years).
 In order to solve the triple system, as well as to further investigate the effect of apsidal motion, an extensive dataset, spanning several decades, is required.

Two stars of spectral type between B3 and B2 in the close orbit, as suggested before by T70, agree with our orbital solution. We estimated  the component masses $M_1 \in [7.6,9.7] M_{\odot}$ and $M_2 \in [6.4;8.2] M_{\odot}$. Both components may be spin-orbit synchronized. Interferometric measurements would be a mayor step foreward in accurately determining  the physical parameters of the system. The triple system $\varepsilon\,$Lup is a very suitable target for large interferometers.

We used the \KOREL disentangling technique as an intermediate step in the study of the two early-B type line-profile variable stars of similar brightness. The variability of primary and secondary seems to be preserved well after disentangling. 

Although the grayscale representations  clearly show signs of the presence of variability, we were not able to detect a dominant frequency in the line profiles of the primary. We thus classify the first component of $\varepsilon\,$Lup as a  $\beta$~Cephei\, suspect.  On the other hand, a candidate pulsational frequency was found in the
variability of the secondary, near 10.36 c d$^{-1}$. Its precise value
could not be determined due to the poor frequency resolution. This
frequency certainly needs confirmation by means of additional
intensive monitoring. Nevertheless, we propose the secondary of $\varepsilon\,$Lup as a new $\beta$~Cephei\, variable. 

Further investigation of the intrinsic variability of both components of $\varepsilon\,$Lup requires a follow-up multi-site campaign with intensive monitoring of
the star, preferably at phases near elongation, by means of
high-resolution spectroscopy. For an example of the benefit of such a
multi-site campaign, we refer to Handler et al.\ (2004) and Aerts et
al.\ (2004b).

The detection of several observational cases of tidally induced oscillations is necessary to
improve our understanding of the relation between tidal interaction
and the excitation  of pulsation modes, and the seismic modelling of such stars.  Effects such as the deformation of the spherical shape of a
star due to tidal forces are usually not taken into account in theoretical calculations. Aerts et al.\ (2002) have shown that, in the case of the
$\delta$\,~Scuti star \object{XX\,Pyx}, deformation of the star due to the
tide-generating potential is clearly more important for 
interpreting  the pulsational behaviour than is the  rotational deformation. The
tide-generating potential is proportional to  $\epsilon_T =
(R_1/a)^3(M_2/M_1)$, while the deformation due to
centrifugal forces is proportional to $\epsilon_C = (f_{\rm
rot}/f_{\rm puls})^2$. The values derived by Aerts et al.\ (2002) for XX\,Pyx were $\epsilon_T = 6
\times 10^{-3}$ and $\epsilon_C = 5 \times 10^{-4}$. We also calculated
the contributions of the tidal forces and rotation on the deviation of
spherical symmetry for the secondary of $\varepsilon\,$Lup.  
Assuming  $i_{\rm orb} = 20.5^{\circ}$, $R_1 \sim 4.7 R_{\odot}$ (Table~\ref{elements}) and assuming spin-orbit synchronization $f_{\rm rot} = 0.403$ c d$^{-1}$, we
derived $\epsilon_T \sim 0.004$ and $\epsilon_C \simeq 0.002$; hence, $\epsilon_T$ is larger than $\epsilon_C$.  As a comparison we 
calculated $\epsilon_C$ and $\epsilon_T$ for the hybrid $\gamma\,$Doradus
star \object{HD\,209\,295} for which tidally induced oscillations indeed have
been detected (Handler et al.\ 2002). As not all parameters are given
by these authors, we  can only give a rough estimate: the value of
$\epsilon_T$ is on the order of $3 \times 10^{-3}$, while  $\epsilon_C$
is 3 orders of magnitude larger. For this system, the rotational effects on the oscillations are thus  larger than  the tidal effects.
These examples show that variety of behaviour exists amongst
close binary systems. Only confrontation between several observational findings
and theory can bring us closer to  understanding the mutual
interaction between  the mechanisms that play a role in the properties of oscillations in close binary systems. 

\begin{acknowledgements}
The authors acknowledge  
 Dr.~P.~Hadrava for sharing his codes \KOREL and \FOTEL.  This study has benefited greatly from the senior
fellowship awarded to PH by the Research Council of the University of Leuven,
which allowed his three-month stay at this university. KU is supported by the Fund for Scientific
Research -- Flanders (FWO) under project G.0178.02 and CA  by
the Research Fund K.U.Leuven under grant GOA/2003/04. 
\end{acknowledgements}

{}

\onecolumn
{\small 
\begin{center}
\begin{longtable}{llllll}
\caption{RVs of the primary and secondary of $\varepsilon\,$Lup. The columns give HJD, the RV of the primary, and an  error estimate, the RV of the secondary and an error estimate, and a label indicating the source of the RV value. Error estimates of the RVs are only available for the  recent data. As error  estimate, we adopted the rms errors of the mean of the RVs calculated by CC from the \SiIII, \MgII, and \HeI\ profiles mentioned in the text.}\label{onlinedata}\\
\hline\hline
Date (HJD) & RV1 & error1 & RV2 & error2 & Note\\ \noalign{\smallskip}
\hline
\endfirsthead 
\caption{continued}\\
\hline\hline
Date (HJD) & RV1 & error1 & RV2 & error2 & Note\\\noalign{\smallskip}
\hline
\endhead
\hline
\endfoot
2417691.7678&  -37.6&       &  62.5&      & A\\
2417703.5991&   31.2&       & -30.0&      & A\\
2418760.9114&   74.9&       & -77.3&      & A\\
2418825.7282&   14.2&       &      &      & A\\
2418829.4971&   52.3&       & -70.6&      & A\\
2419112.9000&    9.2&       &      &      & A\\
2419156.7000&   11.3&       &      &      & A\\
2419157.6720&   74.8&       & -41.5&      & A\\
2419157.8070&   52.5&       & -74.2&      & A\\
2419221.5900&   47.6&       & -73.4&      & A\\
2435297.9640&   28.0&       &  -9.0&      & B \\
2436793.8760&    4.0&       & -19.0&      & B\\
2436798.8560&    8.0&       & -21.0&      & B\\
2438456.5951&  -42.0&       &  54.0&      & C\\
2438463.6087&   77.0&       & -74.0&      & C\\
2438478.6309&    5.0&       &      &      & C\\
2438481.5881&   77.0&       & -66.0&      & C\\
2438494.5070&    8.0&       &      &      & C\\
2438505.4036&    9.0&       &      &      & C\\
2438518.4670&   70.0&       & -67.0&      & C\\
2438520.4831&  -28.0&       &  55.0&      & C\\
2438563.4078&   53.0&       & -44.0&      & C\\
2438588.3454&  -12.0&       &  46.0&      & C\\
2438595.3089&   50.0&       & -40.0&      & C\\
2438598.1947&  -33.0&       &  52.0&      & C\\
2438599.2336&    3.0&       &      &      & C\\
2438600.2796&   73.0&       & -81.0&      & C\\
2438601.2175&    8.0&       &      &      & C\\
2438602.2524&  -21.0&       &  53.0&      & C\\
2438604.1792&   10.0&       &      &      & C\\
2438884.4871&    3.0&       &      &      & C\\
2438906.4533&   42.0&       & -29.0&      & C\\
2439363.1908&    4.0&       &      &      & C\\
2439363.2438&    4.0&       &      &      & C\\
2439625.4602&   10.0&       &      &      & C\\
2439626.4213&   73.0&       & -87.0&      & C\\
2439671.2942&   27.0&       & -47.0&      & C\\
2439696.2405&   -3.0&       &      &      & C\\
2439696.3665&   -3.0&       &      &      & C\\
2439723.1773&    3.0&       &      &      & C\\
2439723.2742&    3.0&       &      &      & C\\
2439756.1914&  -33.0&       &  60.0&      & C\\
2440012.3611&  -31.0&       &  54.0&      & C\\
2440333.3966&   58.0&       & -51.0&      & C\\
2440369.3663&   59.0&       & -59.0&      & C\\
2449965.6037&  -38.4&   2.0 &  47.7&   4.5& D\\
2450195.8947&   76.0&   2.1 & -66.1&   5.5& D\\
2450196.6035&   13.0&   3.5 &  13.7&   6.5& D\\
2452770.5066&   -1.3&   8.2 &      &      & D\\
2452770.5109&   -2.9&   7.7 &      &      & D\\
2452770.5170&   -2.0&   7.6 &      &      & D\\
2452770.5368&   -1.9&   7.3 &      &      & D\\
2452770.5587&   -2.4&   6.8 &      &      & D \\
2452770.6025&   -1.6&   6.2 &      &      & D\\
2452770.6388&   -1.1&   4.4 &      &      & D\\
2452770.6604&   -0.3&   4.2 &      &      & D\\
2452770.6930&   -1.5&   4.2 &      &      & D\\
2452770.7516&   -0.1&   4.6 &      &      & D\\
2452770.7702&   -0.9&   2.2 &      &      & D\\
2452770.7896&   -1.2&   2.3 &      &      & D\\
2452770.8095&    0.0&   2.4 &      &      & D\\
2452770.8296&    0.3&   2.6 &      &      & D\\
2452770.8493&    0.7&   2.7 &      &      & D\\
2452770.8704&    1.3&   2.7 &      &      & D\\
2452770.8899&    0.5&   3.3 &      &      & D\\
2452770.9119&    1.5&   2.8 &      &      & D\\
2452771.5107&   46.7&   4.8 & -47.7&   5.7& D\\
2452771.5300&   48.7&   4.7 & -50.3&   7.5& D\\
2452771.5491&   50.5&   5.5 & -52.2&   8.6& D\\
2452771.5918&   53.3&   5.2 & -55.6&  11.3& D\\
2452771.6262&   55.4&   5.2 & -59.9&  17.4& D\\
2452771.6631&   59.0&   4.6 & -60.6&   7.8& D\\
2452771.7543&   64.6&   6.7 & -70.6&  11.6& D\\
2452771.7782&   66.3&   2.0 & -71.5&  12.4& D\\
2452771.8008&   66.8&   3.3 & -71.2&   7.1& D\\
2452771.8251&   67.8&   3.5 & -74.5&   7.8& D\\
2452771.8491&   69.2&   3.7 & -76.6&  11.7& D\\
2452771.8720&   70.6&   4.5 & -76.1&  13.0& D\\
2452771.8961&   70.5&   2.8 & -79.0&  13.3& D\\
2452772.6170&   11.7&   5.5 &      &      & D\\
2452772.6505&    9.8&   3.4 &      &      & D\\
2452772.7490&    5.0&   3.8 &      &      & D\\
2452772.7740&    4.3&   3.7 &      &      & D\\
2452772.8111&    2.1&   6.4 &      &      & D\\
2452772.8455&    0.9&   5.4 &      &      & D\\
2452773.6156&  -31.9&   4.8 &  46.6&   9.1& D\\
2452773.6476&  -32.5&   4.8 &  46.4&   9.1& D\\
2452773.6813&  -33.8&   4.6 &  46.3&   9.6& D\\
2452773.7155&  -34.6&   3.2 &  48.7&   9.6& D\\
2452773.7450&  -35.7&   3.1 &  48.7&   9.3& D\\
2452773.7722&  -36.2&   3.2 &  48.6&   9.1& D\\
2452773.7945&  -36.8&   3.3 &  48.3&   7.4& D\\
2452773.8105&  -36.0&   4.3 &  50.9&   7.2& D\\
2452773.8389&  -36.5&   4.5 &  49.8&   7.4& D\\
2452773.8674&  -36.7&   4.7 &  50.6&  11.3& D\\
2452774.5575&  -30.9&   4.1 &  48.0&  14.7& D\\
2452774.5683&  -33.9&   2.1 &  41.5&   8.4& D\\
2452775.5329&    3.6&   3.8 &      &      & D\\
2452775.5555&    3.7&   3.7 &      &      & D\\
2452775.5991&    5.2&   3.8 &      &      & D\\
2452775.7024&    6.9&   2.6 &      &      & D\\
2452775.7524&    8.3&   4.9 &      &      & D\\
2452775.8086&   10.2&  13.4 &      &      & D\\
2452775.8306&   11.5&  15.2 &      &      & D\\
2452775.8556&   11.8&  15.8 &      &      & D\\
2452780.2814&    1.5&   5.7 &      &      & E\\
2452780.3087&    1.8&  14.9 &      &      & E\\
2452780.3416&    2.2&   5.5 &      &      & E\\
2452780.3824&    5.6&   8.8 &      &      & E\\
2452780.5423&   32.8&   7.1 & -36.8&  14.4& E\\
2452780.5751&   34.8&   8.6 & -42.9&   5.7& E\\
2452781.3053&   61.3&   5.8 & -70.3&  11.0& E\\
2452781.3111&   60.4&   2.2 & -70.6&   5.9& E\\
2452781.3397&   56.3&   5.2 & -67.3&  13.3& E\\
2452781.3733&   54.7&   5.3 &      &      & E\\
2452781.3930&   53.4&   6.0 &      &      & E\\
2452781.4285&   50.0&   7.9 & -57.7&   8.9& E\\
2452781.4345&   51.1&   4.5 & -57.2&   8.2& E\\
2452781.4598&   48.8&   4.6 & -55.8&   7.3& E\\
2452781.4953&   43.2&   4.6 & -51.5&   4.9& E\\
2452781.5226&   40.4&   5.1 & -48.5&   3.0& E\\
2452781.5506&   38.5&   3.4 & -46.1&   4.1& E\\
2452781.5776&   33.8&   4.3 & -40.5&   8.9& E\\
2452781.6056&   32.6&   4.0 & -34.5&  12.1& E\\
2452783.2466&  -38.7&   4.8 &      &      & E\\
2452783.2527&  -38.7&   4.6 &      &      & E\\
2452783.2601&  -36.9&   3.7 &  48.4&  13.8& E\\
2452783.2666&  -37.3&   3.6 &  46.9&  10.6& E\\
2452783.2745&  -36.8&   2.7 &  45.7&   8.9& E\\
2452783.2815&  -36.9&   2.6 &  45.7&   9.3& E\\
2452783.3250&  -36.1&   3.2 &  46.4&   3.9& E\\
2452783.3745&  -36.3&   2.9 &  46.6&  11.2& E\\
2452783.4047&  -35.6&   3.2 &  46.2&   4.8& E\\
2452783.4123&  -35.9&   3.0 &  44.7&   6.5& E\\
2452783.4423&  -35.4&   1.3 &  45.1&   7.3& E\\
2452783.4911&  -35.2&   1.8 &  44.5&   7.1& E\\
2452783.4974&  -35.6&   2.4 &  40.4&   5.9& E\\
2452783.5244&  -35.6&   1.8 &  41.2&   6.2& E\\
2452783.5491&  -35.3&   1.2 &  41.7&   7.0& E\\
2452783.5988&  -34.0&   6.0 &  41.1&  13.9& E\\
2452784.2934&   -5.7&   5.4 &  -6.8&   8.4& E\\
2452784.3026&   -5.3&   4.7 &  -6.4&   8.2& E\\
2452784.4391&   -5.0&   5.4 &  -2.1&   4.1& E\\
2452784.4463&   -4.8&   4.9 &  -1.5&   3.3& E\\
2452784.4661&   -0.6&   4.2 &  -2.6&   4.3& E\\
2452784.4708&   -2.2&   5.1 &  -2.4&   4.5& E\\
2452784.4787&   -5.5&   5.4 &  -3.2&   3.7& E\\
2452784.5072&   -4.5&   4.7 &  -4.2&   5.0& E\\
2452784.5116&   -4.5&   4.9 &  -3.0&   4.6& E\\
2452784.5481&   -4.3&   5.1 &  -2.7&   5.5& E\\
2452784.5542&   -2.4&   3.5 &  -2.6&   4.7& E\\
2452784.6059&   -1.9&   4.1 &  -1.8&   5.1& E\\ \noalign{\smallskip}\hline
\multicolumn{6}{l}{\scriptsize{A: Campbell \& Moore (1928); B: Buscombe \& Morris (1960); C: Thackeray (1970); }} \\
\multicolumn{6}{l}{\scriptsize{D: Average of the first moments of the \SiIII~4553~\AA\ and 4568~\AA\ profiles; }} \\
\multicolumn{6}{l}{\scriptsize{E: Median of the cross-correlation measurements calculated from the \SiIII~4568~\AA,}}\\
\multicolumn{6}{l}{\scriptsize{\MgII~4481~\AA, \HeI~5016~\AA, \HeI~4917~\AA, \HeI~5876~\AA\ and \HeI~6678~\AA\ profiles}} 
\end{longtable}
\end{center}
}
\end{document}